 \documentclass[final,1p,times]{elsarticle}

\usepackage{listings}
\usepackage{enumitem}
\usepackage{multirow}
\usepackage{rotating}
\usepackage{epstopdf}
\usepackage{graphicx}
\usepackage{subfigure}

\usepackage{times}

\usepackage{makeidx}
\usepackage{multirow}
\usepackage{multicol}
\usepackage[dvipsnames,svgnames,table]{xcolor}
\usepackage{graphicx}
\usepackage{epstopdf}
\usepackage{ulem}

\usepackage{array}
\usepackage{amsmath,amssymb,amsfonts}
\usepackage{algorithm}
\usepackage{algpseudocode}
\usepackage{graphicx}
\usepackage{textcomp}

\usepackage{epsfig}
\usepackage{makecell}
\usepackage{longtable,booktabs}

\usepackage[colorlinks,
linkcolor=black,
anchorcolor=blue,
citecolor=green
]{hyperref}

\usepackage{longtable}

\journal{Computers \& Mathematics with Applications}

\begin{document}
\begin{frontmatter}
	
	
		
	\title{	juSFEM: A Julia-based Open-source Package of Parallel Smoothed Finite Element Method (S-FEM) for Elastic Problems}

\author[label1]{Zenan Huo}
\author[label1]{Gang Mei\corref{cor1}}
\ead{gang.mei@cugb.edu.cn}
\cortext[cor1]{Corresponding authors}
\author[label1]{Nengxiong Xu\corref{cor1}}
\ead{xunengxiong@cugb.edu.cn}

\address[label1]{School of Engineering and Technology, China University of Geosciences (Beijing), 100083, Beijing, China}

\begin{abstract}

The Smoothed Finite Element Method (S-FEM) proposed by Liu G.R. can achieve more accurate results than the conventional FEM. Currently, much commercial software and many open-source packages have been developed to analyze various science and engineering problems using the FEM. However, there is little work focusing on designing and developing software or packages for the S-FEM. In this paper, we design and implement an open-source package of the parallel S-FEM for elastic problems by utilizing the Julia language on multi-core CPU. The Julia language is a fast, easy-to-use, and open-source programming language that was originally designed for high-performance computing. We term our package as juSFEM. To the best of the authors’ knowledge, juSFEM is the first package of parallel S-FEM developed with the Julia language. To verify the correctness and evaluate the efficiency of juSFEM, two groups of benchmark tests are conducted. The benchmark results show that (1) juSFEM can achieve accurate results when compared to commercial FEM software ABAQUS, and (2) juSFEM only requires 543 seconds to calculate the displacements of a 3D elastic cantilever beam model which is composed of approximately 2 million tetrahedral elements, while in contrast the commercial FEM software needs 930 seconds for the same calculation model; (3) the parallel juSFEM executed on the 24-core CPU is approximately 20$\times$ faster than the corresponding serial version. Moreover, the structure and function of juSFEM are easily modularized, and the code in juSFEM is clear and readable, which is convenient for further development.

\end{abstract}

\end{frontmatter}


\subsection*{List of Abbreviations }
\begin{table*}[htbp]

		\begin{tabular}{p{60pt}p{180pt}}
			BLAS& Basic Linear Algebra Subprograms  \\
			COO& COOrdinate  \\
			CSC& Compressed Sparse Column  \\
			CSR& Compressed Sparse Row  \\
			CPU& Central Processing Unit  \\			
            FEM & Finite Element Method  \\
			GPU& Graphics Processing Unit  \\
			LAPACK& Linear Algebra PACKage  \\
			MKL& Intel® Math Kernel Library  \\
			S-FEM& Smoothed Finite Element Method  \\
			T3& Three-noded Triangular Element  \\ 
			T4& Four-noded Tetrahedral Element  \\
			T10& Ten-noded Terrahedral Element  \\	
		\end{tabular}

\end{table*}

\newpage

\section{Introduction}
\label{sec1}

With the rapid development of computer technology, the Finite Element Method (FEM) has been widely used to solve science and engineering problems \cite{roy2008recent,xu2016tetrahedral}. The FEM is a mesh-based numerical method that needs to divide the study domain into a valid computational mesh consisting of interconnected and non-overlapping elements. In the FEM, it is necessary to first create a shape function of the displacement field, then calculate the strain field by exploiting the relationship between displacement and strain, and finally create and solve the system of discrete equations to obtain the nodal displacements, strain, and stress. 

After years of development, the conventional FEM become quite mature, and is capable of accurately solving various problems in science and engineering. However, the FEM also encounters several problems \cite{leveque2002finite}. For example, when modeling and analyzing the large deformation \cite{he2010coupled} or dynamic crack propagation, the FEM may have difficulties, and the precision of the numerical results is often unsatisfactory. Moreover, when analyzing the three-dimensional solid problems, elements such as tetrahedrons are not continuous on the contact faces, which leads to the stiffness being too rigid. In addition, volume locking also occurs when the material is incompressible or nearly incompressible \cite{wihler2006locking}. 

To deal with the aforementioned weakness of the conventional FEM, in recent years, Liu G.R. et al \cite{liu2016smoothed} proposed a series of S-FEMs by combined the FEM with the smooth strain technology \cite{yoo2004stabilized}. The S-FEM is between the FEM and the pure Meshfree method. Compared with the FEM, the S-FEM requires a modification or reconstruction of the strain field based on background mesh. The background mesh can be exactly the same as the computational mesh in FEM. When using the same number of elements as the FEM, the S-FEM reduces the mesh quality requirements while providing more accurate (smoothed) computational results. 

Although the S-FEM can achieve more accurate numerical results than the FEM, it is less efficient than the FEM. In the S-FEM, the interaction between elements makes the bandwidth of the global stiffness matrix larger than that of the global stiffness matrix in the FEM \cite{nguyen2009face}. Moreover, smoothing domains need to be created and smoothing strain should be calculated. Therefore, the S-FEM is computationally more expensive than the conventional FEM. When to analyze large-scale problems in science or engineering, the computational efficiency of the S-FEM should be improved. 

There are two typical strategies to improve the computational efficiency of the S-FEM. One is to theoretical redesign the algorithm structure in combination with the characteristics of the S-FEM to improve the computational efficiency. The other is to utilize the parallelism on multi-core CPU or many-core GPU. As a variation of the S-FEM, the edge-based smooth element method (ES-FEM) is parallelized on the GPU \cite{cai2018parallel}.

Because the process of the S-FEM is quite similar to the FEM, most current S-FEM programs are modified on the basis of the FEM \cite{phan2013edge}. Reference \cite{li2016automatic} introduced an effective algorithm for establishing the smoothing domains, realizing the automation of 3D entity calculation and adaptive analysis. However, the current S-FEM program, which uses a programming language such as C/C++ or Fortran, usually requires high demand of programming skill, and the code is difficult to read and modify. For programs written in high-level languages such as Python and MATLAB, the code is readable but computationally inefficient \cite{sanner1999python}.

In summary, quite few efforts are dedicated to developing the open-source packages of the S-FEM by comprehensively considering the accuracy, efficiency, readability, and ease-of-development. To balance the program execution efficiency and the ease of implementing the program, in this paper, we design and implement an open-source package of parallel S-FEM for elastic problems by using the Julia language \cite{bezanson2017julia}. We term our package as juSFEM. To the best of the authors’ knowledge, juSFEM is the first package of parallel S-FEM developed with the Julia language.

The Julia language is a fast, easy-to-use, and open-source programming language that was originally designed for high-performance computing. In combination with the principle of the S-FEM and the characteristics of parallel computing, we redesign the algorithm structure and develop a package of the S-FEM, juSFEM, to solve the 3D elastic problems. 

Our contributions in this work can be summarized as follows.

(1) A Julia-based parallel package of the S-FEM for elastic problems is developed.

(2) The redesigned strategy in juSFEM significantly improves the computing efficiency.

(3) The structure and function of juSFEM are modularized, and the code is clear and easy to understand, which is convenient for subsequent improvement.

The rest of this paper is organized as follows. Section 2 presents the background introduction to the S-FEM and the Julia language. Section 3 introduces the implementation details of the package juSFEM. Section 4 provides several examples to validate the correctness and evaluate the efficiency of juSFEM. Section 5 analyzes the performance, advantages, and shortcomings of the package juSFEM. Finally, Section 6 concludes this work.

\section{Background}
\label{sec2}

In this section, we will present brief introduction to (1) the S-FEM and (2) the Julia language.

\subsection{Smoothed Finite Element Method (S-FEM)}

The S-FEM is proposed by Liu G.R. \cite{liu2016smoothed} to overcome the weakness of the conventional FEM such as the overly stiff phenomenon \cite{zienkiewicz1971reduced}. Because of its excellent performance, S-FEM has been widely applied in material mechanics \cite{liu2015smoothed}, fracture mechanics \cite{chen2010singular,chen2011novel}, acoustics \cite{liu2009node,li2014analysis}, heat transfer \cite{li2014smoothed,cui2016steady}, and fluid–structure interactions \cite{he2011coupled}. The S-FEM has been proved to be an effective and robust numerical method for various types of problems in the past decade \cite{zeng2018smoothed}.

The essential idea behind the S-FEM is to modify the compatible strain field or construct a strain field using only the displacements. Such modification and construction are performed within the so-called smoothing domain. The method of modification is only suitable for three-noded triangular element (T3) or four-noded tetrahedral element (T4) since their compatible strain field is easy to obtain. In order to support multiple element types, we choose the method of constructing strain field in juSFEM. 

Depending on the different methods of construction of the smooth domains, the S-FEM can be divided into four categories: the node-based smoothed FEM (NS-FEM) \cite{liu2009node}, the edge-based smoothed FEM (ES-FEM) \cite{liu2009edge}, the face-based smoothed FEM (FS-FEM) \cite{nguyen2009face}, and the cell-based smoothed FEM (CS-FEM) \cite{liu2007smoothed}. The calculation process of the FS-FEM using the T4 element is as follows.

\textbf{STEP 1. Discretization  of the 3D problem domain}

The FS-FEM generally uses the T4 element to discrete three-dimensional problem domains. When the T4 element is used, mesh generation is performed in the same manner as in the FEM, for example, by employing the widely used Delaunay tetrahedralization algorithm \cite{shewchuk2016delaunay}.

\textbf{STEP 2. Creating a displacement field}

\begin{figure}[htbp]
	\centering
	\includegraphics[width=0.9\textwidth]{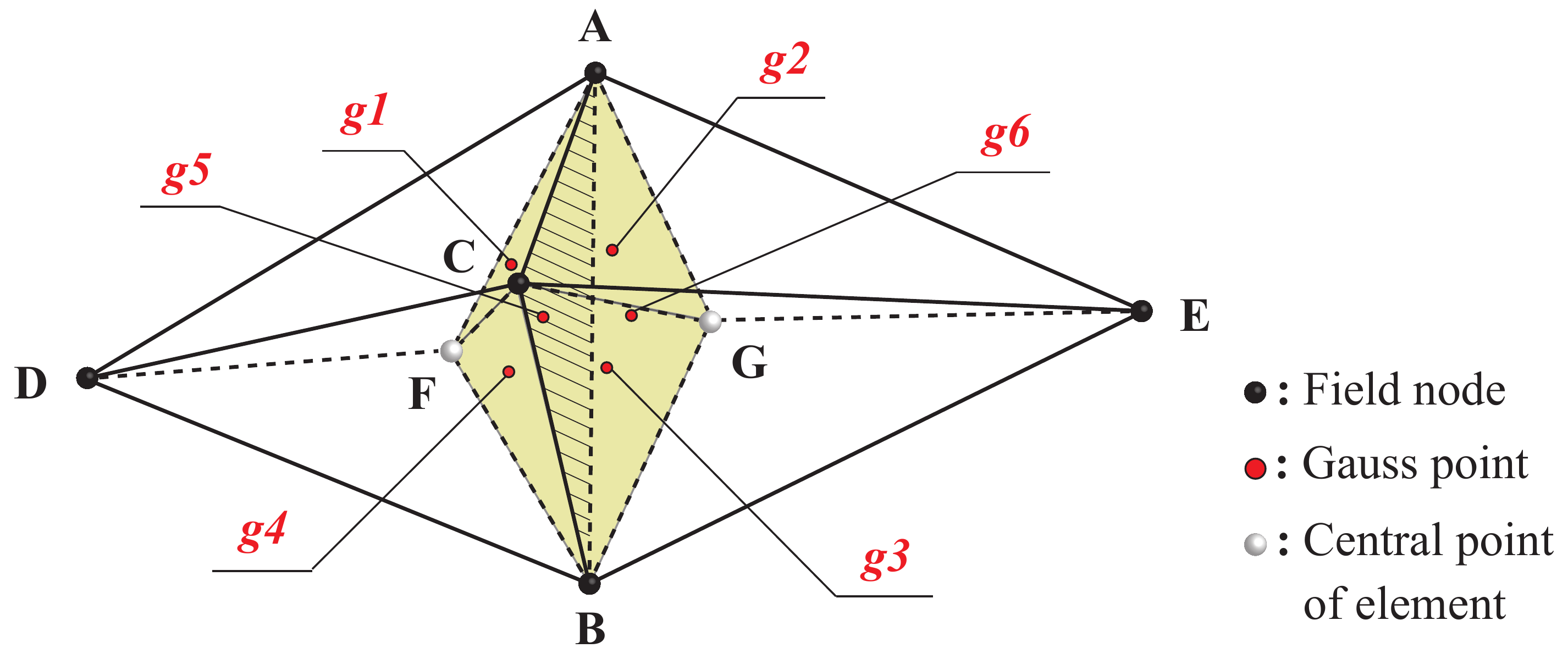}
	\caption{Two adjacent tetrahedral elements and the smoothing domain (shaded domain) formed based on the interface $\triangle ABC$} \label{fig1}
\end{figure}

As shown in Figure \ref{fig1}, the attributes of the face are divided into two types: the interior (triangular face $\triangle ABC$) and the exterior (triangular face $\triangle ADC$, $\triangle ACE$, etc.). For the interior face, the body center points F and G are calculated in the tetrahedra ABCD and ABCE. Point F connects with three points in $\triangle ABC$ to form a new tetrahedron ABCF. Point G connects with three points in $\triangle ABC$ to form a new tetrahedron ABCG. The hexahedron ABCFG formed by two tetrahedra is the smoothing domain of the interior face $\triangle ABC$. The external face $\triangle ABD$ has no adjacent tetrahedral elements, and the body center point F connects with three points of $\triangle ACD$ to form a new tetrahedron ACDF, which is the smoothing domain generated by the face $\triangle ACD$.

The displacement field function is then assumed as:
\begin{equation}
\label{eq1}
\overline{\mathbf{u}}(\mathbf{x})=\sum\limits_{I=1}^{{{N}_{n}}}{{{\mathbf{N}}_{I}}}(\mathbf{x}){{\overline{\mathbf{d}}}_{I}}=\underbrace{\left\{ {{N}_{1}}(x)\ {{N}_{1}}(x)\ \ldots \ {{N}_{{{N}_{n}}}}(x) \right\}}_{N}\underbrace{\left\{ \begin{matrix}
	\overline{{{d}_{1}}}  \\
	\overline{{{d}_{2}}}  \\
	\vdots   \\
	\overline{{{d}_{{{N}_{n}}}}}  \\
	\end{matrix} \right\}}_{\overline{d}}=N(x)\overline{d}
\end{equation}

\textbf{STEP 3. Evaluation of the compatible strain field}

The FS-FEM uses the boundary integral method to calculate the node shape function value at the relevant point according to the node information of the smoothing domains and the connection information between the elements. In the FS-FEM, the smoothing strain is calculated as:

\begin{equation}
\label{eq2}
\overline{\varepsilon}\left(x_{k}\right)=\int_{\Omega_{k}^{\mathrm{s}}} \tilde{\varepsilon}(x) \phi\left(x_{k}-x\right) \mathrm{d} \Omega
\end{equation}
where $\overline{\varepsilon}\left(x_{k}\right)$ is the compatible strain field evaluated using an assumed displacement field, $\Omega_{k}^{\mathrm{s}}$ is the smoothing domain, and $\phi\left(x_{k}-x\right)$ is the smoothing function:

\begin{equation}
\label{eq3}
\phi \left( {{x}_{k}}-x \right)=\left\{ \begin{array}{*{35}{l}}
1/V_{k}^{s}, & x\in \Omega _{k}^{s}  \\
0, & x\notin \Omega _{k}^{s}  \\
\end{array} \right.
\end{equation}
where $V_{k}^{s}$ is the volume of the smoothing domain. 

Using Green’s divergence theorem, the smoothing strain matrix can be solved by the boundary integral of the smoothing domains:

\begin{equation}
\label{eq4}
\overline{\mathbf{B}}_{I}\left( {{\mathbf{x}}_{k}} \right)=\left( \frac{1}{V_{k}^{\text{s}}} \right)\int_{{{\Gamma }^{k}}}{{{\mathbf{n}}^{k}}}{{\mathbf{N}}_{I}}(\mathbf{x})\text{d}\Gamma =\left[ \begin{matrix}
{{\overline{b}}_{Ix}}({{x}_{k}}) & 0 & 0  \\
0 & {{\overline{b}}_{Iy}}({{x}_{k}}) & 0  \\
0 & 0 & {{\overline{b}}_{Iz}}({{x}_{k}})  \\
0 & {{\overline{b}}_{Iz}}({{x}_{k}}) & {{\overline{b}}_{Iy}}({{x}_{k}})  \\
{{\overline{b}}_{Iz}}({{x}_{k}}) & 0 & {{\overline{b}}_{Ix}}({{x}_{k}})  \\
{{\overline{b}}_{Iy}}({{x}_{k}}) & {{\overline{b}}_{Ix}}({{x}_{k}}) & 0  \\
\end{matrix} \right]
\end{equation}

In Eq. \eqref{eq4}:

\begin{equation}
\label{eq5}
\overline{b}_{Ih}\left( {{x}_{k}} \right)=\left( \frac{1}{V_{k}^{s}} \right)\int_{{{\Gamma }^{k}}}{{{N}_{I}}}(x)n_{h}^{k}(x)d\Gamma ,\ \ \ (h=x,y,z)
\end{equation}
where $\Gamma^{k}$ is the face boundary of the smoothing domain $\Omega_{k}^{s}$ , and in Figure \ref{fig1}, it is the six triangular faces of the hexahedron ABCFG. $n_{h}^{k}(x)$ is the unit normal direction vector of each face, and its form is:

\begin{equation}
\label{eq6}
\boldsymbol{n}^{k}(\boldsymbol{x})=\left[\begin{array}{ccc}{n_{x}^{k}} & {0} & {0} \\ {0} & {n_{y}^{k}} & {0} \\ {0} & {0} & {n_{z}^{k}} \\ {0} & {n_{z}^{k}} & {n_{y}^{k}} \\ {n_{z}^{k}} & {0} & {n_{x}^{k}} \\ {n_{y}^{k}} & {n_{x}^{k}} & {0}\end{array}\right]
\end{equation}

When using the strain field that is linearly compatible with the face boundary, and the integration is conducted on the face. Therefore, Eq. \eqref{eq5} can be simplified to:

\begin{equation}
\label{eq7}
\overline{b}_{I h}\left(x_{k}\right)=\left(\frac{1}{V_{k}^{\mathrm{s}}}\right) \sum_{i=1}^{P} n_{i h}^{k} N_{I}\left(x_{i}^{\mathrm{G}}\right) A_{i}^{k}, \quad(h=x, y, z)
\end{equation}
where $P$ is the number of face boundaries $\Gamma^{k}$ ; for the interior, $\Gamma^{k}=6$ , and for the exterior, $\Gamma^{k}=4$ .   $x_{i}^{\mathrm{G}}$is the Gauss point of $\Gamma^{k}$ ,  $A_{i}^{k}$ is the area of $\Gamma^{k}$ , and $n_{i h}^{k}$ is the outer unit normal vector.

The value of the shape function of Gauss points (centroid points) on the boundary of the smoothing domains is listed in Table \ref{tab1}.

\begin{table}[!h]
	\caption{Shape function values at different sites on the smoothing domain boundary associated with the face $\triangle ABC$ in Figure \ref{fig1}}
	\centering
		\begin{tabular}{|c|c|c|c|c|c|c|}
		\hline
		\textbf{Site} & \textbf{Node A} & \textbf{Node B} & \textbf{Node C} & \textbf{Node D} & \textbf{Node E} & \textbf{Description} \\ \hline
		\textbf{A}    & 1          & 0          & 0          & 0          & 0          & Field node           \\ \hline
		\textbf{B}    & 0          & 1          & 0          & 0          & 0          & Field node           \\ \hline
		\textbf{C}    & 0          & 0          & 1          & 0          & 0          & Field node           \\ \hline
		\textbf{D}    & 0          & 0          & 0          & 1          & 0          & Field node           \\ \hline
		\textbf{E}    & 0          & 0          & 0          & 0          & 1          & Field node           \\ \hline
		\textbf{F}    & 1/4        & 1/4        & 1/4        & 1/4        & 0          & Centroid of element  \\ \hline
		\textbf{G}    & 1/4        & 1/4        & 1/4        & 0          & 1/4        & Centroid of element  \\ \hline
		\textbf{g1}   & 5/12       & 1/12       & 5/12       & 1/12       & 0          & \textbf{Gauss point} \\ \hline
		\textbf{g2}   & 5/12       & 1/12       & 5/12       & 0          & 1/12       & \textbf{Gauss point} \\ \hline
		\textbf{g3}   & 1/12       & 5/12       & 5/12       & 0          & 1/12       & \textbf{Gauss point} \\ \hline
		\textbf{g4}   & 1/12       & 5/12       & 5/12       & 1/12       & 0          & \textbf{Gauss point} \\ \hline
		\textbf{g5}   & 5/12       & 5/12       & 1/12       & 1/12       & 0          & \textbf{Gauss point} \\ \hline
		\textbf{g6}   & 5/12       & 5/12       & 1/12       & 0          & 1/12       & \textbf{Gauss point} \\ \hline
		\end{tabular}
		\label{tab1}
\end{table}

\textbf{STEP 4. Establish the system equation}

In the FS-FEM, it uses the smoothed Galerkin weak form and the assumed displacement and smoothing strain fields to establish the discrete linear algebraic system of equations instead of using the Galerkin weak form used in the FEM. In this process, the FS-FEM requires only a simple summation over all of the smoothing domains. The linear system of equations when using T4 element is as follow:

\begin{equation}
\label{eq8}
\overline{K}^{FS-FEM}\overline{d}=f
\end{equation}
where $\overline{K}^{FS-FEM}$ is the smoothed stiffness matrix whose entries are given by:

\begin{equation}
\label{eq9}
\overline{\mathbf{K}}_{I J}^{\mathrm{FS}-\mathrm{FEM}}=\int_{\Omega} \overline{\mathbf{B}}_{I}^{T} \mathbf{c} \overline{\mathbf{B}}_{J} \mathrm{d} \Omega=\sum_{k=1}^{N_{f}} \int_{\Omega_{k}^{s}} \overline{\mathbf{B}}_{I}^{T} \mathbf{c} \overline{\mathbf{B}}_{J} \mathrm{d} \Omega=\sum_{k=1}^{N_{f}} \overline{\mathbf{B}}_{I}^{T} \mathbf{c} \overline{\mathbf{B}}_{J} V_{k^{\prime}}^{s}
\end{equation}

\textbf{STEP 5. Impose essential boundary conditions}

In the S-FEM, the essential boundary conditions are satisfied when the displacement field is assumed. Therefore, the shape functions constructed and used for creating the displacement field have the important $Delta$ function properties. This practical approach of treating the essential boundary conditions is exactly the same as in FEM: essentially by removing (or modifying) the rows and columns of the stiffness matrix.

\begin{figure}[H]
	\centering
	\includegraphics[width=0.7\textwidth]{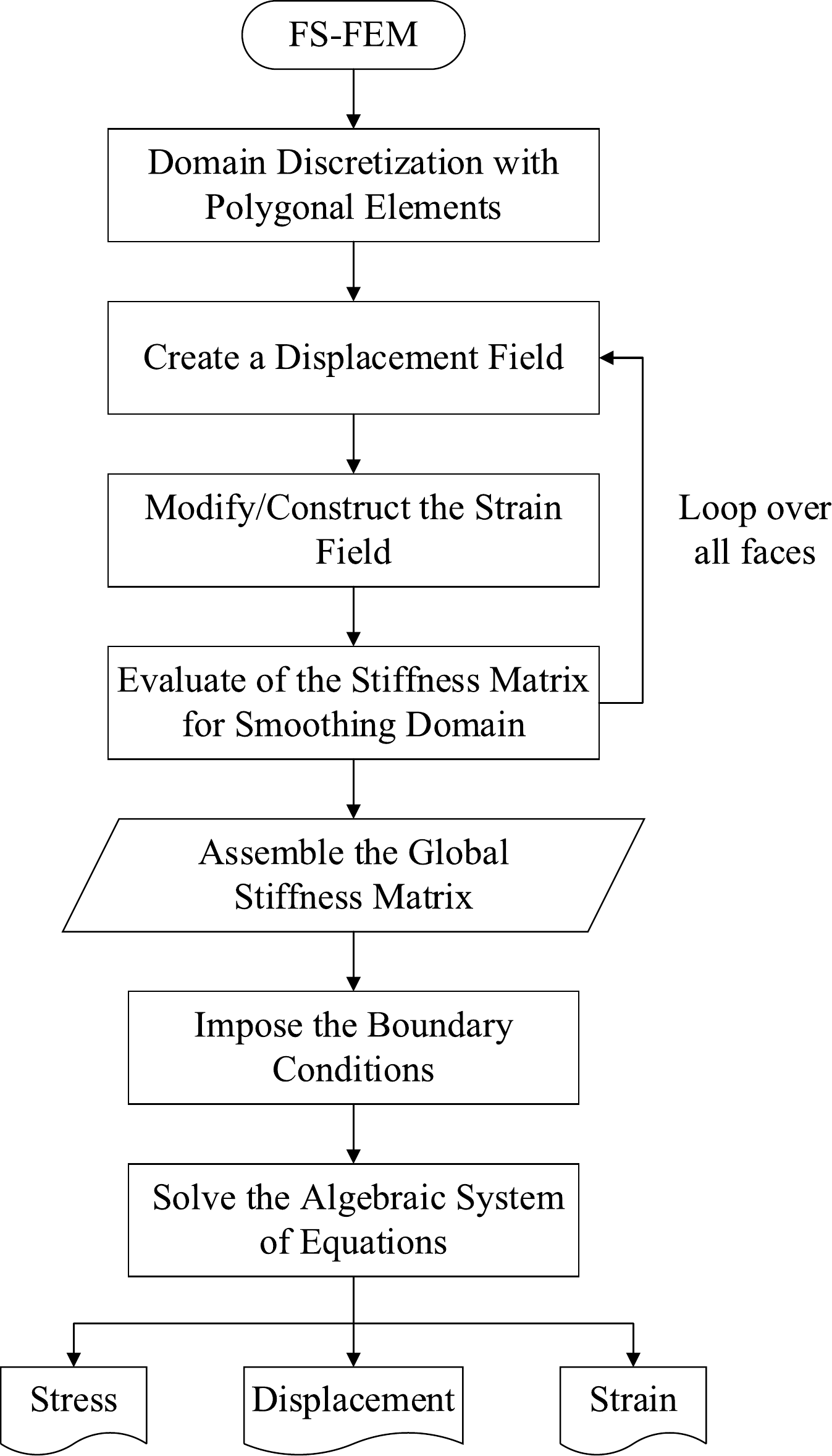}
	\caption{Flow chart of the FS-FEM} \label{fig2}
\end{figure}

The overall calculation process of the FS-FEM is illustrated in Figure \ref{fig2}.

\subsection{Julia Language}

\begin{figure}[!h]
	\centering
	\includegraphics[width=1.1\textwidth]{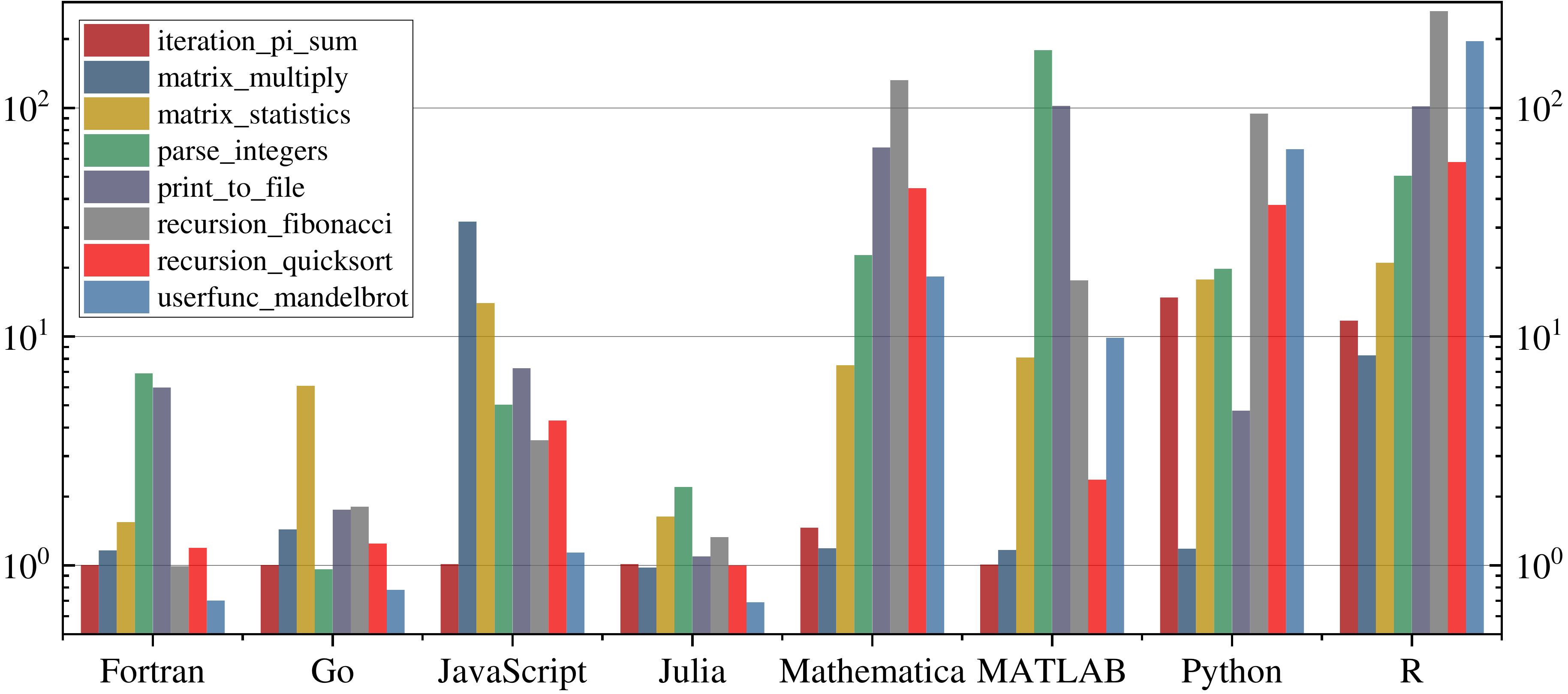}
	\caption{Julia benchmarks (the benchmark data shown above were computed with Julia v1.0.0, Go 1.9, JavaScript V8 6.2.414.54, MATLAB R2018a, Anaconda Python 3.6.3, and R 3.5.0. C and Fortran are compiled with GCC 7.3.1, taking the best timing from all optimization levels. C performance is 1.0, smaller is better, the benchmark data is download from \cite{benchmarks})} \label{fig3}
\end{figure}

The Julia language is a fast, easy-to-use, and open-source development scripting language under the MIT license \cite{julia}. It was originally designed to meet the needs of high-performance numerical analysis and computational science. Fortran is the programming language to formally adopt high-level programming concepts, and the goal was to write high-level, general-purpose formulas that could be automatically converted into efficient code. Not only does the complexity of Fortran require high programming ability of programmers, but the written code is also less readable. In recent years, dynamic languages such as Python and MATLAB have gradually become popular \cite{pine2019introduction}, but their computational efficiency is limited. Julia created a new numerical computing method that is a dynamic language based on the parallelization and distributed computing of a mathematical programming language. Julia's performance is close to that of C/C++ and other statically compiled languages, making it ideal for writing efficient S-FEM programs; see Figure \ref{fig3}. A package written in Julia is more readable and easier for users to extend and improve.

Julia language has been used in many practical engineering problems. Frondelius et al. \cite{frondelius2017juliafem} proposed a FEM framework in Julia language, which allows the use of simple programming models for distributed processing of large finite element models across computer clusters. Otter et al. \cite{otter2019thermodynamic} described a prototype that demonstrates how thermodynamic property and thermo-fluid pipe component modeling enhanced via Julia language. To implement large-scale image processing algorithms, Lage-Freitas et al. \cite{7729158} proposed a solution which transparently processes distributed underlying computing resources with the advantages of Julia DistributedArrays, and they provided a digital programming interface to process huge data sets.

There are three levels of parallel computing in Julia: (1) Julia Coroutines (Green Threading), (2) Multi-Threading (Experimental), (3) Multi-core or Distributed Processing.

Julia Coroutines is suitable for small tasks. Multi-threading in Julia is currently an experimental interface and may cause unknown errors. To achieve program stability and high efficiency, we choose multi-core processing to realize the parallelism of the S-FEM algorithm.

\begin{figure}[H]
	\centering
	\includegraphics[width=0.9\textwidth]{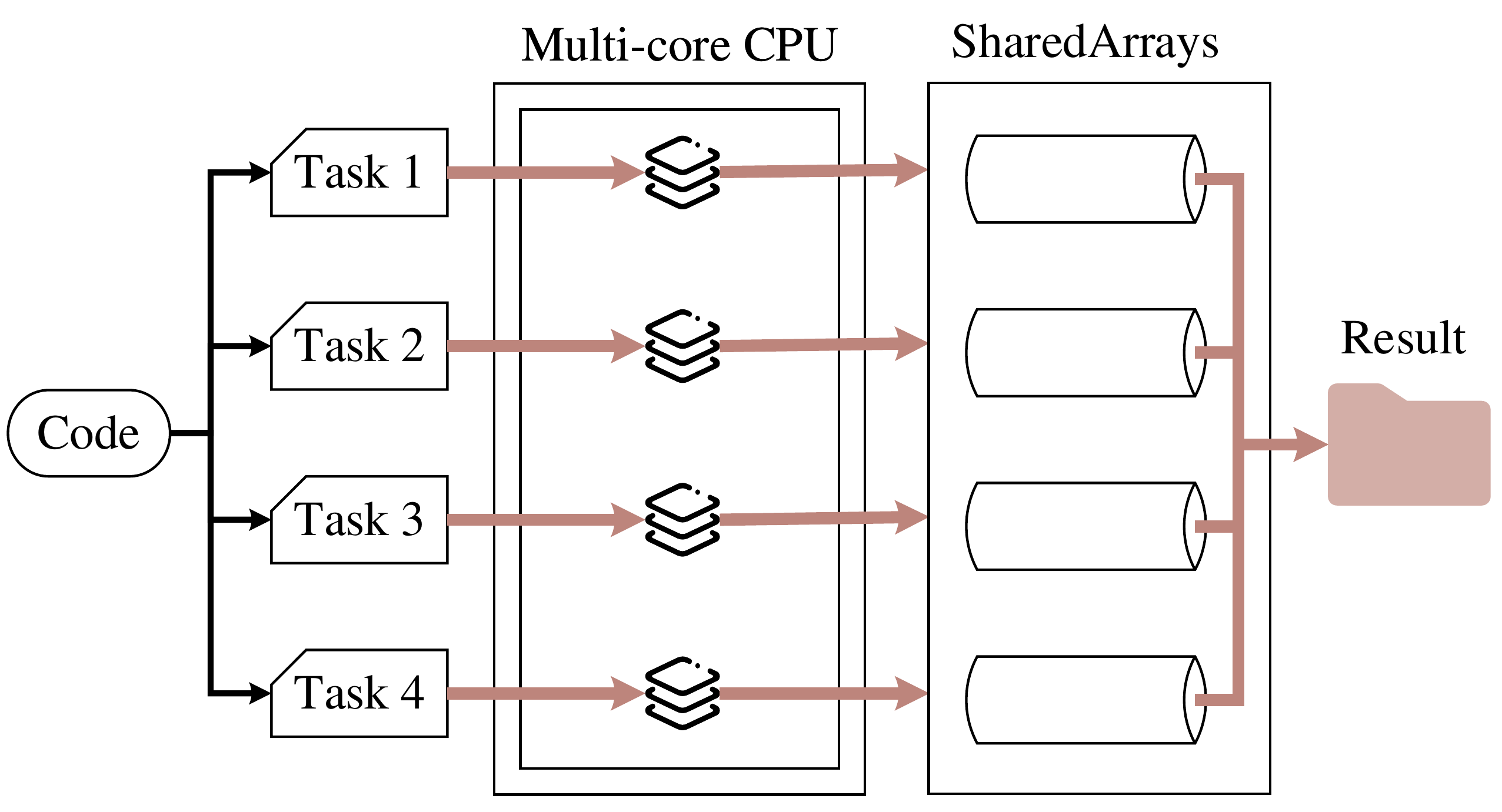}
	\caption{Illustration of the work flow of multi-core computing in Julia} \label{fig4}
\end{figure}

Nowadays, multi-core computing is ubiquitous in ordinary computers, and according to Moore's law, we expect the number of cores per chip to increase as the number of transistors increases \cite{chai2007understanding}. As shown in Figure \ref{fig4}, multi-core computing slices program tasks and redistributes them to each process to start computing. In Julia, the use of "\texttt{SharedArrays}" allows different processes to access common data.

The functions “\texttt{addprocs},”“\texttt{rmprocs},” “\texttt{workers},” and others are available as a programmatic means of adding, removing, and querying the processes. In particular, the iterations do not occur in a specified order, and writes to variables or arrays will not be globally visible because iterations run on different processes. Any variables used inside the parallel loop will be copied and broadcasted to each process. “\texttt{SharedArrays}” can be used to deal with situations where multiple processes are operating in an array at the same time. 

\section{The Developed Package juSFEM}
\label{sec3}

\subsection{Overview}

In this paper, we design and implement an open-source package of parallel S- FEM for elastic problems by utilizing the Julia language. We term our package as juSFEM. To the best of the authors’ knowledge, juSFEM is the first package of parallel S-FEM developed with the Julia language, which is publicly available at 
\url{https://github.com/Elliothuo/juSFEM}.

\begin{figure}[H]
	\centering
	\includegraphics[width=1.0\textwidth]{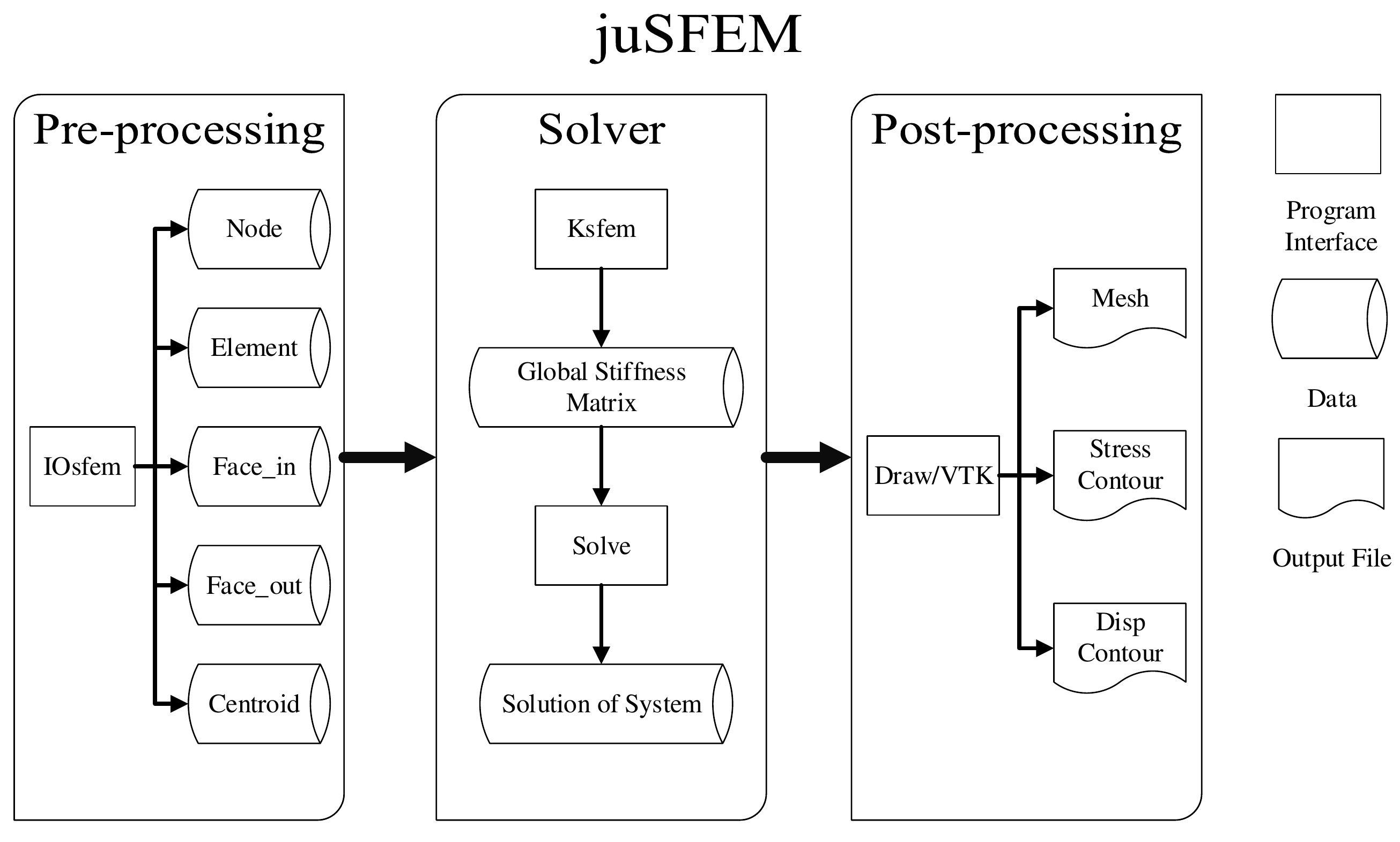}
	\caption{Structure of the package juSFEM} \label{fig5}
\end{figure}

The package juSFEM is composed of three components: pre-processing, solver, and post-processing; see the structure of juSFEM in Figure \ref{fig5}.

(1) \textbf{Pre-processing}: the background mesh is achieved using mature mesh generation algorithms, and a new form for indexing mesh topological information is proposed to prepare for the building of smoothing domain.

(2) \textbf{Solver}: the global stiffness matrix is assembled according to the S-FEM principle; the boundary conditions are imposed, and the system of linear equations is solved to obtain nodal displacements.

(3) \textbf{Post-processing}: the numerical results are plotted interactively.

\subsection{Pre-processing}

We used TetGen \cite{si2015tetgen} to generate a high-quality tetrahedral mesh. The tetrahedral mesh is stored in “\texttt{node}”, “\texttt{face}” and “\texttt{ele}” files. On this basis, we designed five pre-processing matrices to record subdivision information according to the computational characteristics of the FS-FEM. We further processed the mesh detail by combining the FS-FEM and parallel computing features, and finally used five matrices to store the mesh detail correspondingly.

\begin{figure}[htbp]
	\centering
	\includegraphics[width=1\textwidth]{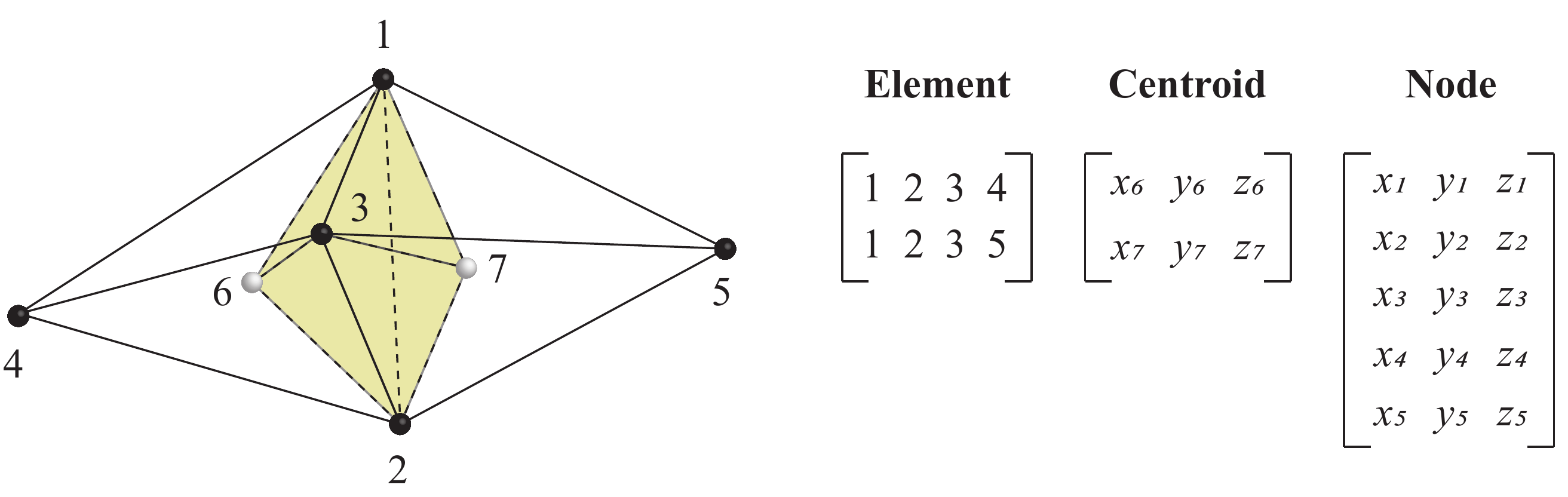}
	\caption{Illustration of the matrices “\texttt{Node}”, “\texttt{Element}”, “\texttt{Centroid}” (Using two adjacent tetrahedrons as example)} \label{fig6}
\end{figure}

First, the “\texttt{node}” file is read into the matrix “\texttt{Node}” in the Julia program. “\texttt{ele}” is stored in the matrix “\texttt{Element}”. The number of rows in the “\texttt{Node}” matrix is the number of nodes. Each row has three entries, i.e., the x, y and z coordinates. The row number of the “\texttt{Element}” matrix is the number of all tetrahedral elements in the mesh. There are four columns, which are the four points that make up the tetrahedron. A new matrix, “\texttt{Centroid}”, is created with the same number of rows as “\texttt{Element}” and three columns. Through the four points in each line of “\texttt{Element}”, the body center of the tetrahedron is obtained, and the x, y and z coordinates of the body center are stored in each line of “\texttt{Centroid}”, respectively; see Figure \ref{fig6}.

\begin{figure}[H]
	\centering
	\includegraphics[width=1.1\textwidth]{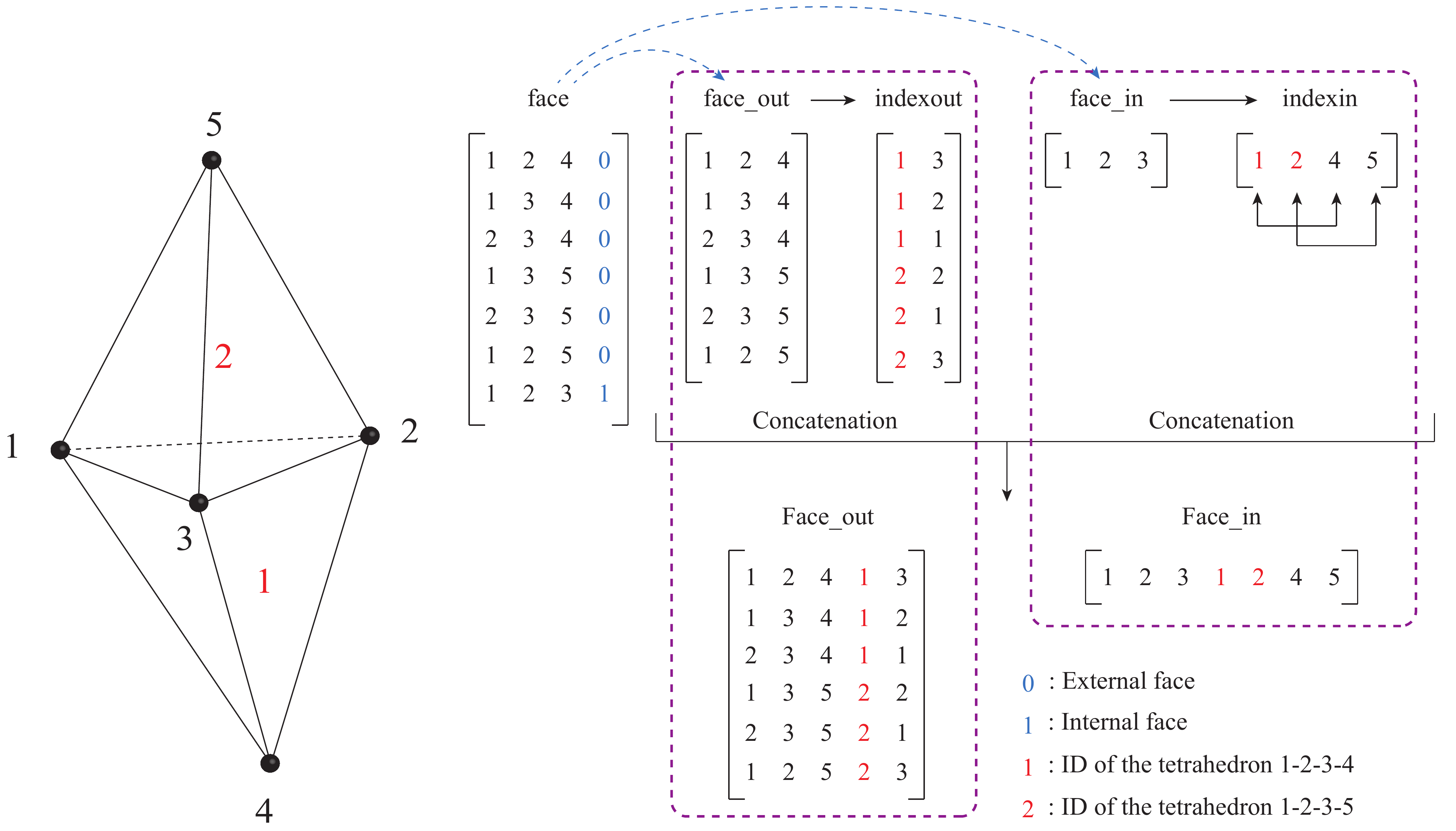}
	\caption{Illustration of processing “\texttt{face}” file  (Using adjacent tetrahedrons as an example)} \label{fig7}
\end{figure}

In the FS-FEM, the smoothing domain is built based on faces. The number of rows in the “\texttt{face}” file is the number of all faces in the mesh model. The first three columns of each row are the indices of points composing of the triangular faces. As shown in Figure \ref{fig7}, if the fourth column is "\texttt{1}", then the triangle face is inside; if it is "\texttt{0}", then the triangle face is outside. The fourth column is calculated by TetGen, which is used to determine the properties of the current triangle face. We divide all the faces of the model into two parts based on the values of the fourth column; the external faces are stored in the “\texttt{face\_out}” matrix, and the internal faces are stored in the “\texttt{face\_in}” matrix, both of which have three columns. Next, we create two new matrices named “\texttt{indexout}” and “\texttt{indexin}”. The matrix “\texttt{indexout}” has the same number of rows as “\texttt{face\_out}” with 2 columns, and the matrix “\texttt{indexin}” has the same number of rows as “\texttt{face\_in}” with 4 columns. By looping over the “\texttt{face\_out}” and “\texttt{face\_in}” matrices, it can determine which tetrahedron each face belongs to. 

For the matrix “\texttt{indexout}”, the sequence number of the tetrahedron is added to the first column, and the remaining point of the tetrahedron is added to the second column. Because “\texttt{face\_in}” faces belong to two tetrahedra, the first two columns of “\texttt{indexin}” are the IDs of tetrahedra, and the last two columns are the IDs of the remaining points in the tetrahedron. Finally, we use the “\texttt{Concatenation}” function in Julia to merge (1) “\texttt{face\_out}” and “ \texttt{indexout}” and (2) “\texttt{face\_in}” and “\texttt{indexin}”. Finally, we obtain the matrix “\texttt{face\_out}” with five columns and “\texttt{face\_in}” with seven columns. The above operations are illustrated in Figure \ref{fig7}. Note that in the mesh generation package, TetGen, the index starts at “\texttt{0}”, while in Julia it starts at “\texttt{1}.”

To efficiently process the “\texttt{face}” file, we utilize the parallelism on multi-core CPU in Julia. We choose “\texttt{SharedArrays}” to deal with data interference in multiple processes. “\texttt{SharedArrays}” use shared system memory to map the same array across many processes. “\texttt{SharedArrays}” indexing (assignment and routines) works just as with regular arrays . We set “\texttt{indexin}” and “\texttt{indexout}” to the “\texttt{SharedArrays}” type, which enables different processes to operate at the same time \cite{document}. An example of an external face is listed in Algorithm \ref{alg:1}.

\begin{algorithm}[H]
	\caption{Procedure of processing the external face}
	\label{alg:1}
	\begin{algorithmic}[1]
		\Require
		\texttt{face\_out, Element}
		\Ensure
		\texttt{Face\_out}
		\State Set \texttt{indexout} to \texttt{SharedArrays}.
		\label{code:fram:1}
		\State \textbf{for} \texttt{face\_out} every row:
		
		Determine which tetrahedron the current triangular face belongs to.
		
		Write judgment result in \texttt{indexout}.	
		
		\noindent \textbf{end}		
		\label{code:fram:2}
		\State Merge \texttt{face\_out} and \texttt{indexout} as \texttt{Face\_out}.
		\label{code:fram:3}\\
		\Return \texttt{Face\_out}
	\end{algorithmic}
\end{algorithm}

After pre-processing, we obtain five matrices: “\texttt{Centroid}”, “\texttt{Face\_in}”, “\texttt{Face\_out}”, “\texttt{Element}”, and “\texttt{Node}”.

\subsection{Solver}

The process of the solver is divided into two stages: (1) the assembly of stiffness matrix and (2) the solution of linear equations, both of which utilize the parallelism to improve the efficiency. In the above two stages, there are three critical issues that need to carefully be addressed. 

(1) Large-scale computation often leads to insufficient memory. Thus, commonly it needs to
use the compression of the global stiffness matrix in a compressed format to reduce computer
memory requirements. The conventional compression formats include COO, CSC, and CSR \cite{jain2002parallel}. Multidimensional arrays in Julia are stored in column-major order. This means that the arrays are stacked one column at a time. Therefore, we choose the COO format to store the global stiffness matrix, and we transform it into the CSC format to improve the speed when solving the linear equations.

(2) In parallel computing, one of the critical problems is the data interference in multiple processes. For example, iterations do not occur in a specified order, and writes to variables or arrays will not be globally visible because iterations run on different processes. Any variables used inside the parallel loop will be copied and broadcast to each process. Therefore, it needs to set the global stiffness matrix to “\texttt{SharedArrays}” type and determine the position and corresponding value of filling the stiffness matrix according to the index of the loop so that each process can assign the stiffness matrix concurrently.

(3) After obtaining the linear equations, it needs to efficiently solve the system of linear equations to obtain the nodal displacements. We invoke PARDISO from the Intel® Math Kernel Library (MKL) \cite{mkl} to solve the equations. PARDISO is a parallel direct sparse solver interface.

\begin{figure}[!h]
	\centering
	\includegraphics[width=1.0\textwidth]{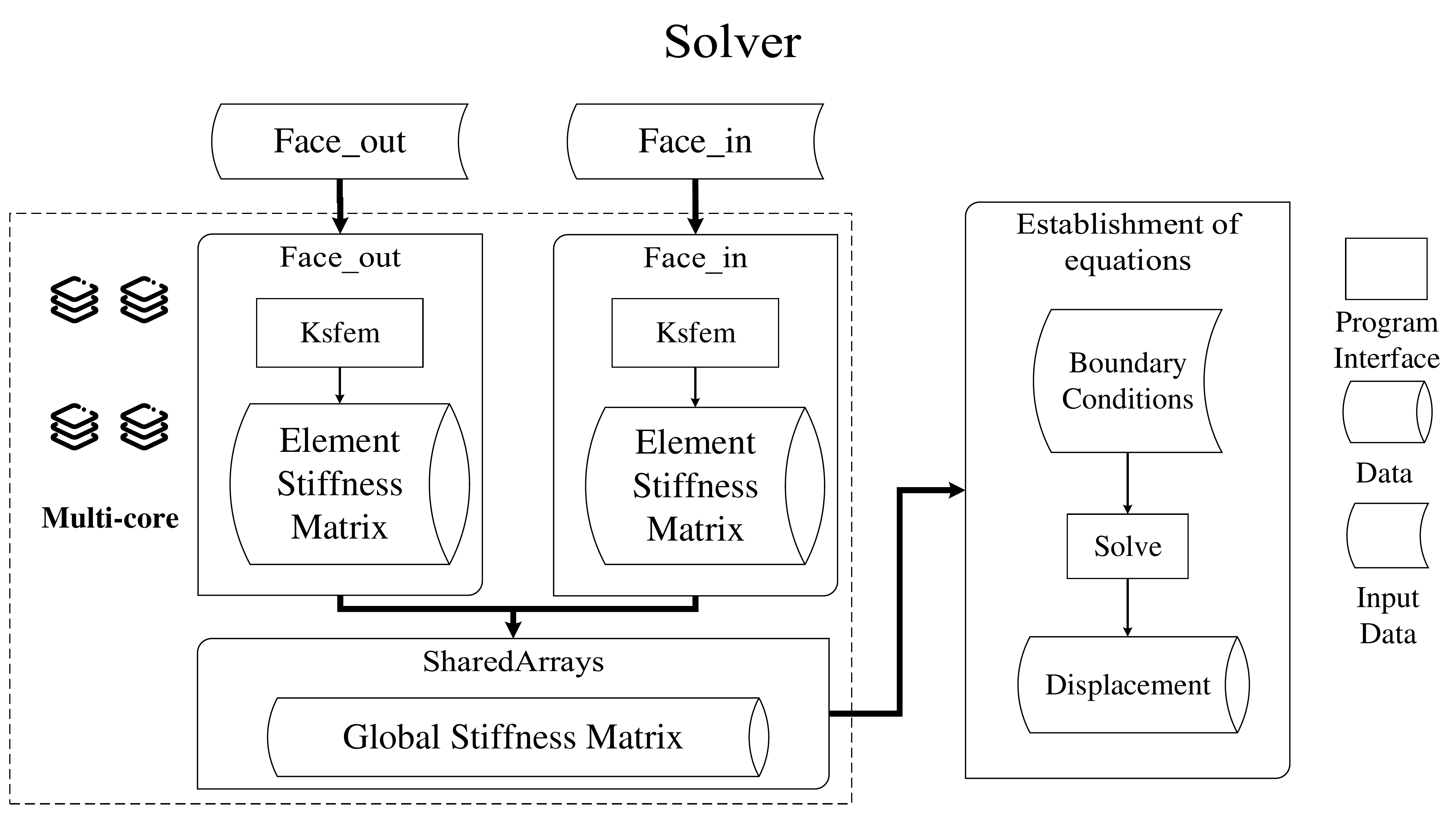}
	\caption{Process of the Solver} \label{fig8}
\end{figure}

The process of the solver in juSFEM is illustrated in Figure \ref{fig8}. After the pre-processing, the global stiffness matrix is first assembled. The assembly of the stiffness matrix is divided into two steps: (1) the assembly for the external faces and (2) the assembly for the internal faces. For the external faces, each smoothing domain is a tetrahedron composed of four points, and each point has three degrees of freedom; thus, the element stiffness matrix for each external face is a $12 \times 12$ matrix. Similarly, the element stiffness matrix for each internal face is a $15 \times 15$ matrix. 

The detailed process of the solver in juSFEM is introduced as follows.

\textbf{STEP 1.}
The calculation of the element stiffness matrices for the internal face and external face are basically the same. First, the volumes of the two tetrahedrons to which the interior face belongs are calculated. This step is implemented in the file “\texttt{volume.jl}”:

\begin{equation}
\label{eq10}
V_{ABCD}=\frac{1}{6}\left|\begin{matrix}x_A&y_A&z_A&1\\x_B&y_B&z_B&1\\x_C&y_C&z_C&1\\x_D&y_D&z_D&1\\\end{matrix}\right|
\end{equation}
where $x_i$, $y_i$, $z_i$ is the coordinate of point $i$.

\textbf{STEP 2.}
The area of each triangular face $S_{\Delta A B C}$ is calculated in Eq. \eqref{eq11}. For the internal face unit, there are six faces. This step is implemented in the file “\texttt{area.jl}”.

\begin{equation}
\label{eq11}
S_{\Delta A B C}=\frac{1}{2}|\overrightarrow{A B} \times \overrightarrow{A C}|
\end{equation}

\begin{figure}[htbp]
	\centering
	\includegraphics[width=0.5\textwidth]{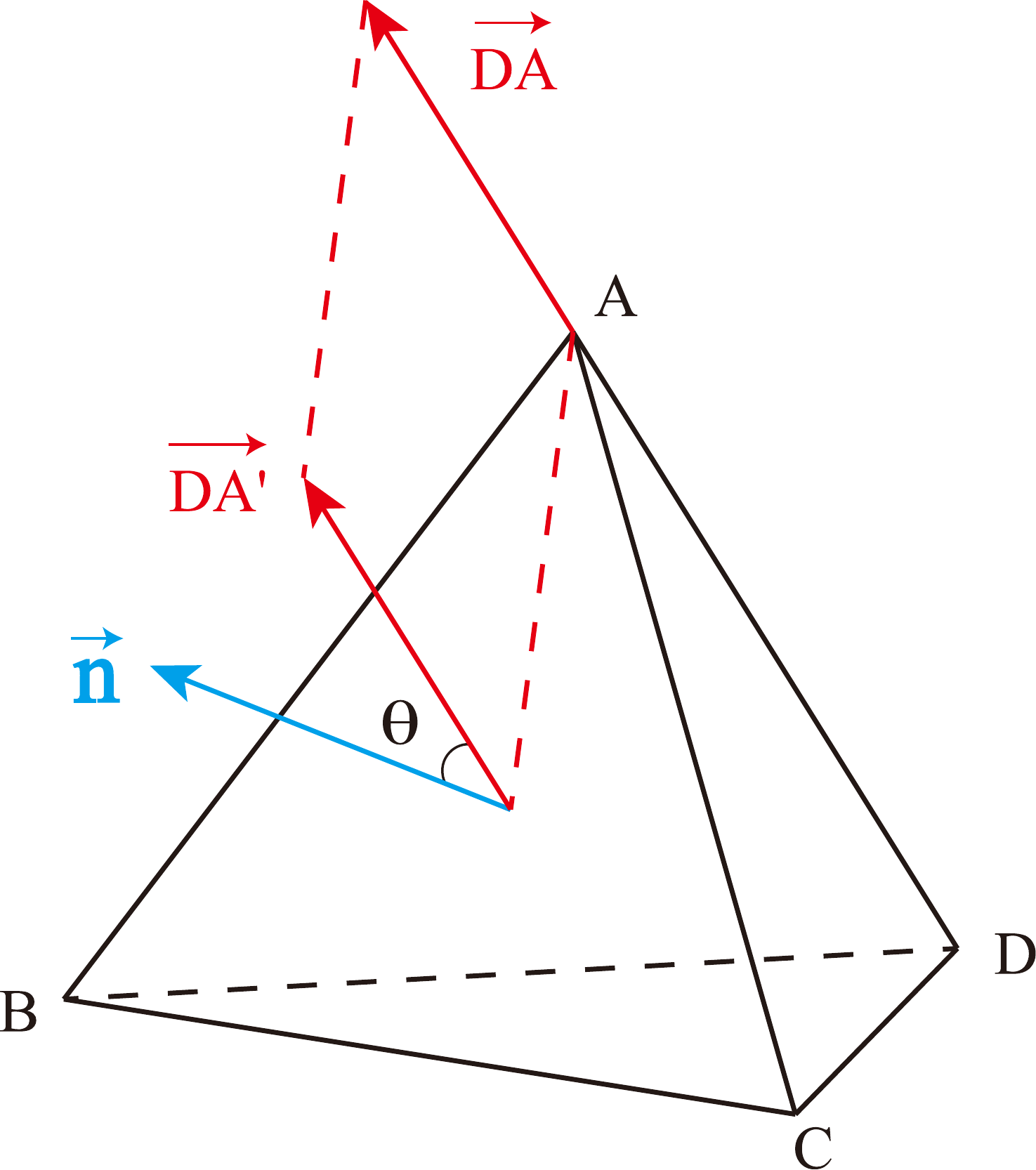}
	\caption{Calculation of the normal vector of an external face} \label{fig9}
\end{figure}

\textbf{STEP 3.}
The normal unit vectors of faces are calculated. As shown in Figure \ref{fig9}, when calculating the normal vector of the first face, it takes the vectors $\vec{AB}$ and $\vec{AC}$ of any two sides in the triangular face and calculate the normal vector $\vec{n}$ of the face through the cross-product of the vector. After that, the element stiffness matrix can be assembled directly according to the principle of the FS-FEM. This step is implemented in the file “\texttt{vectorin.jl}”.

\textbf{STEP 4.}
According to the method by which the sparse matrix is constructed in Julia, we need three one-dimensional arrays. The first array \texttt{IK}, in the order of each row, represents the row number of each value in the global stiffness matrix. The second array \texttt{JK}, in the order of each row, represents the column number of each value in the global stiffness matrix. The third array \texttt{VK}, in the order of each row, represents each value in the global stiffness matrix.

The sparse function can automatically accumulate the values in the same location, thus, the sizes of \texttt{IK}, \texttt{JK} and \texttt{VK} can be determined in advance:

\begin{equation}
\label{eq12}
size\left(IK\right)=size\left(JK\right)=size\left(VK\right)=N_e\times144+N_i\times225
\end{equation}
where $N_e$ is the number of external faces, $N_i$ is the number of internal faces.

Moreover, the arrays \texttt{IK}, \texttt{JK}, and \texttt{VK} are copied into “\texttt{SharedArrays}” in advance. Because the calculation of the stiffness matrix of each element is not data-dependent, no data interference will occur during the parallel calculation of each process.

As mentioned above, we calculate the volume, area and normal vector of the smoothing domains where the inner face is located; and the calculation for each inner face is independent. Therefore, the above calculations are suitable to be parallelized on the multi-core CPU. And the calculated results is also able to save into the three arrays, \texttt{IK}, \texttt{JK}, and \texttt{VK} in parallel. 

The process of assembling the stiffness matrix for the external faces is listed in Algorithm \ref{alg:2}.

\begin{algorithm}[H]
	\caption{Assembly of the global stiffness matrix}
	\label{alg:2}
	\begin{algorithmic}[1]
		\Require
		\texttt{Node, Element, Centroid, Face\_out}
		\Ensure
		\texttt{IK,JK,VK}
		\State  Set \texttt{IK,JK,VK} to \texttt{SharedArrays};
		\label{code:fram:1}		
		\State  \textbf{for} every external surface
			
			Calculate the area of each triangle in the smoothing domain.
		
			Calculate the volume of the smoothing domain.
		
		    Calculate the shape function of Gauss point in smoothing domain.
		
			Calculate the normal unit vector for every triangular faces.
		
			Store results into a compressed fashion.
			
			\noindent \textbf{end}
		\label{code:fram:2}\\
		\Return \texttt{IK,JK,VK}
	\end{algorithmic}
\end{algorithm}

\textbf{STEP 5.}
 After performing similar operations on the external face, \texttt{IK}, \texttt{JK} and \texttt{VK} are completed. The “\texttt{Sparse()}” function is used to construct the stiffness matrix in the CSC format. And the penalty function method is used to impose the boundary conditions. Also, the nodal load is calculated. After that, the MKL is configured, and the \texttt{Pardiso.jl} \cite{pardiso} package is invoked to solve the system of linear equations to first obtain the nodal displacements and then the stain and stress.

\subsection{Post-processing}

In this subsection, we will introduce how to plot the numerical results calculated using the developed package juSFEM.

\subsubsection{Visualization in ParaView}

\begin{figure}[!h]
	\centering
	\includegraphics[width=0.8\textwidth]{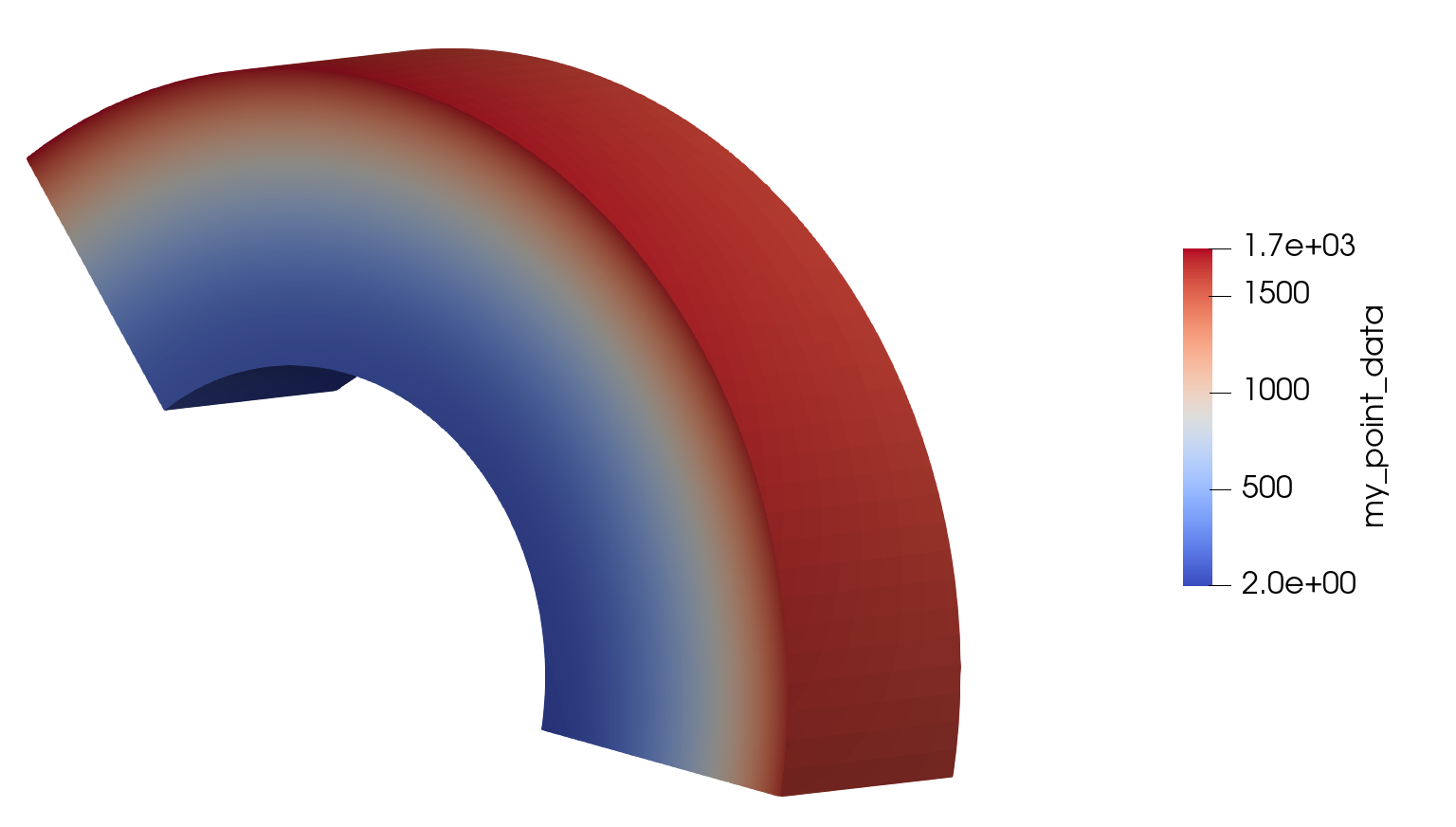}
	\caption{Plotting with the ParaView} \label{fig10}
\end{figure}

After the performing of solver, the nodal displacements can be obtained by solving the system of linear equations, and then the strain and stress can be calculated according to the nodal displacements. To plot these numerically calculated results, we employ the open-source software ParaView \cite{paraview} for the visualization. First, The background mesh and associated displacements, strain, and stress are stored and converted into the “\texttt{vtu}” format file using Julia's open-source library “\texttt{WriteVTK.jl}” \cite{vtk}. Then, the results storing in the “\texttt{vtu}” format file are imported into the ParaView for visualization (Figure \ref{fig10}).

\subsubsection{Visualization in PlotlyJS}

\begin{figure}[H]
	\centering
	\includegraphics[width=0.9\textwidth]{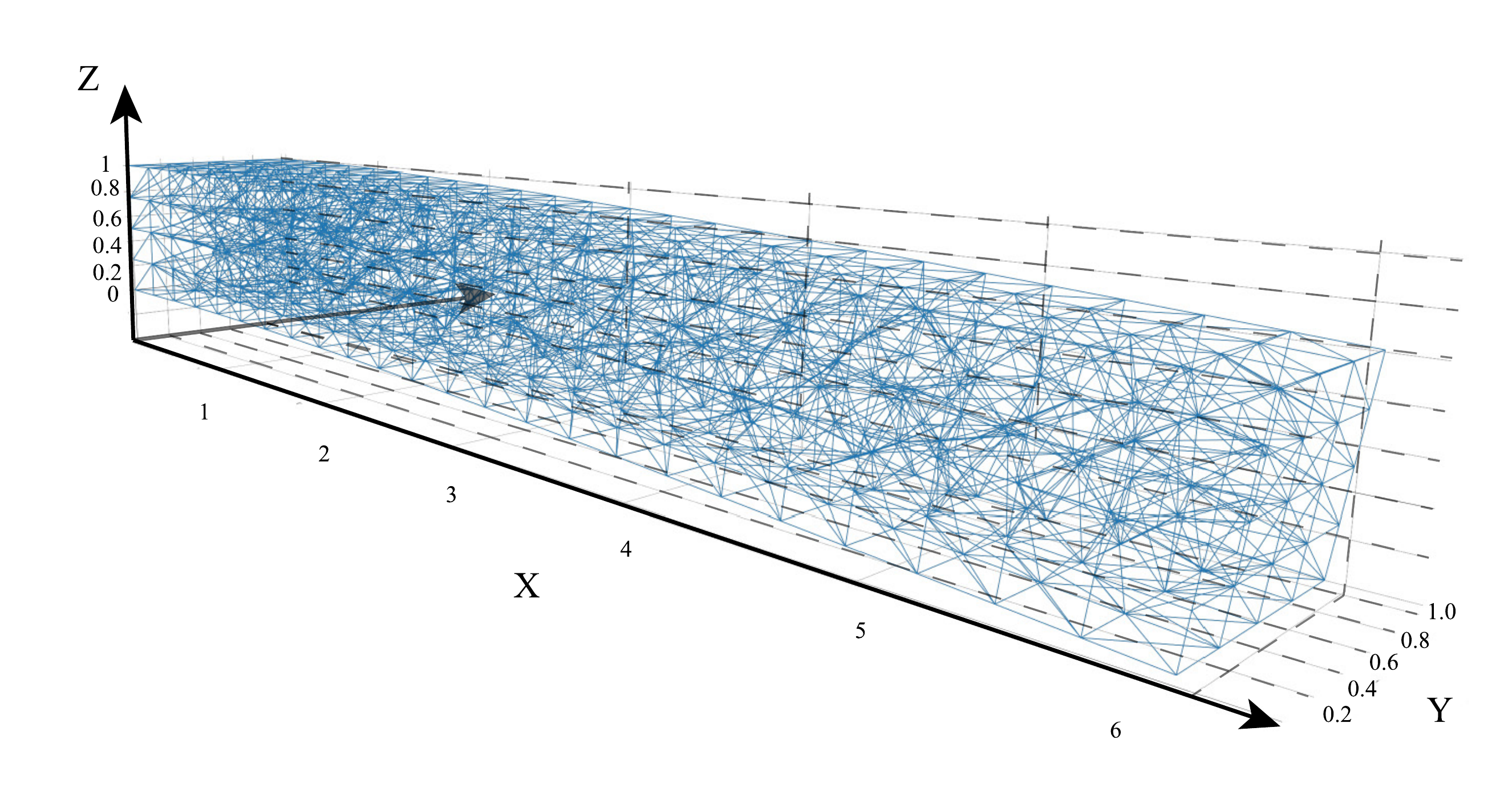}
	\caption{Plotting of the displacement with the PlotlyJS} \label{fig11}
\end{figure}

We also use “\texttt{PlotlyJS.jl}” \cite{plotlyjs} to plot the displacement of the example in an interactive manner. In Julia, we first input the coordinates of each node in the mesh model, the details of each triangular face and the change of displacement of each node, and then plot the numerically calculated results interactively; see Figure \ref{fig11}.

\section{Validation and Evaluation of juSFEM}
\label{sec4}

To evaluate juSFEM's correctness and efficiency, two groups of experiments are conducted on a workstation computer. The specifications of the workstation are listed in Table \ref{tab2}.

\begin{table}[H]
	\caption{Specifications of the employed workstation computer}
	\centering
		\begin{tabular}{|c|c|}
		\hline
\textbf{Specifications}      & \textbf{Details}         \\ \hline
\textbf{OS}                  & Windows 10 Professional  \\ \hline
\textbf{CPU}                 & Intel Xeon 5118          \\ \hline
\textbf{CPU Frequency (GHz)} & 2.30                     \\ \hline
\textbf{CPU Cores}           & 24                       \\ \hline
\textbf{CPU RAM (GB)}        & 128                      \\ \hline
\textbf{Julia Version}       & Julia 1.1.1              \\ \hline         
		\end{tabular}
		\label{tab2}
\end{table}

\subsection{Verification of the Accuracy of juSFEM}

\begin{figure}[!h]
	\centering
	\subfigure[]{
		\label{fig:12a}       
		\includegraphics[width=0.7\textwidth]{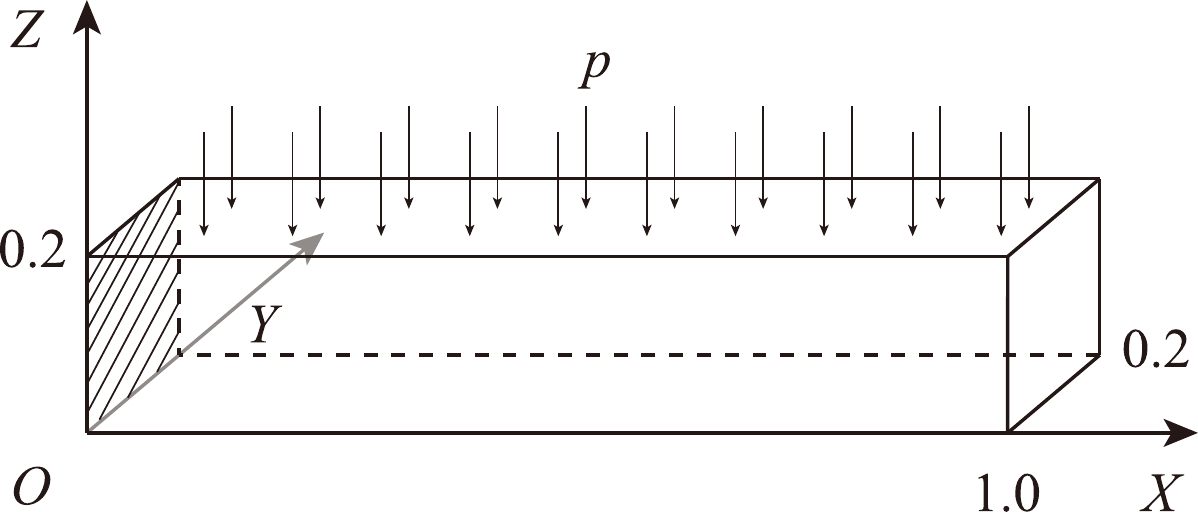}
	}
	\hspace{1em}
	\subfigure[]{
		\label{fig:12b}    
		\includegraphics[width=0.8\textwidth]{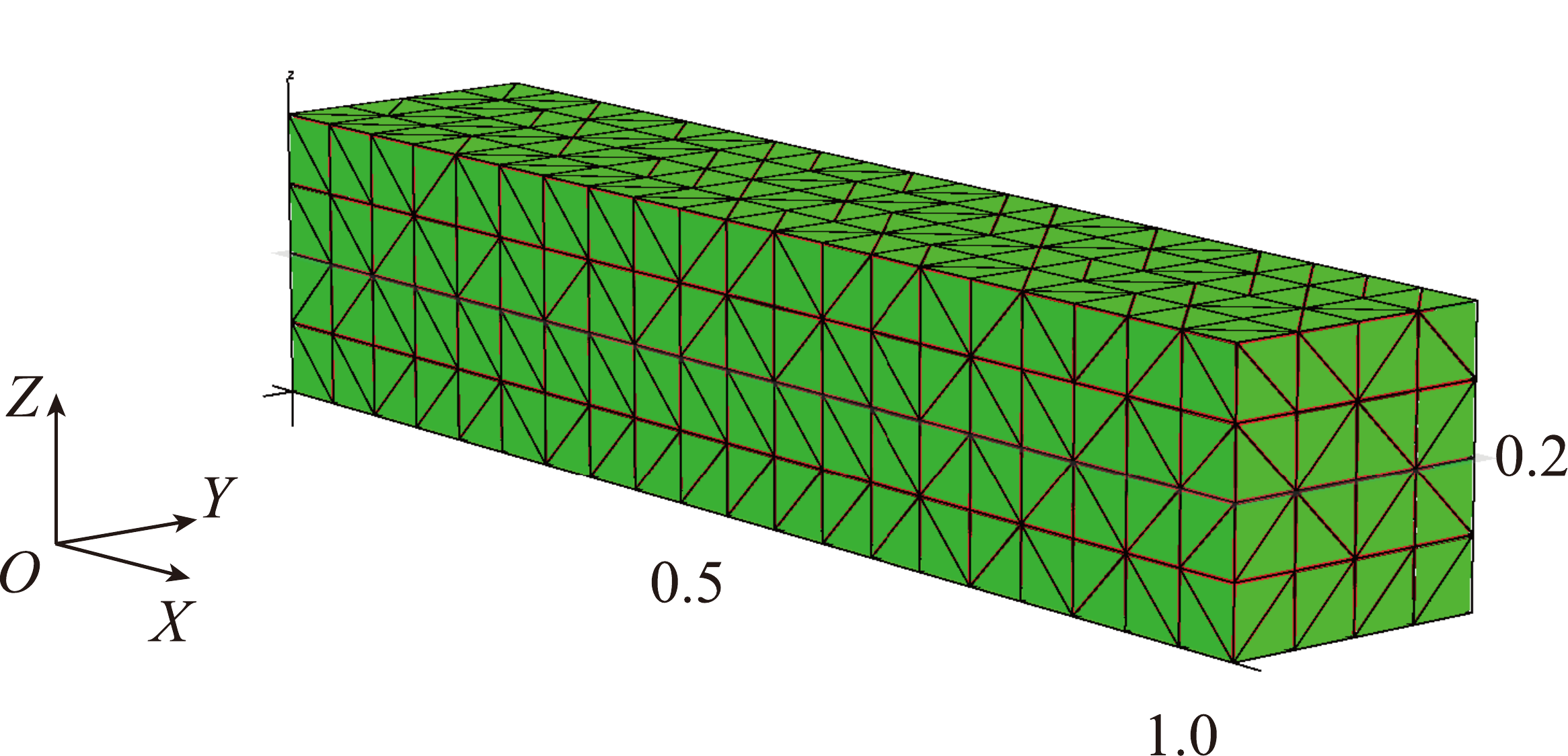}
	}
	\caption{(a) A 3D cantilever beam subjected to a uniform pressure on the top face and (b) a mesh with four-node tetrahedral elements.}
	\label{fig12}       
\end{figure}

To verify the correctness of juSFEM, a 3D elastic cantilever beam model is analyzed to compare the computational accuracy. In this example, the size of the cantilever beam is  $1\times0.2\times0.2(m)$ ; see Figure \ref{fig12}. The upper face of the cantilever beam is subjected to a uniform pressure $p=12500(N/m^2)$ in the direction of $–Z$. The elastic modulus $E=2E8(N/m^2)$ and Poisson's ratio $v=0.3$. The model consists of 458 nodes and 1,576 tetrahedral elements. 

To verify the calculation accuracy, the displacements of the 3D elastic cantilever beam model is calculated and compared in the following three scenes.

\begin{figure}[!h]
	\centering
	\includegraphics[width=0.9\textwidth]{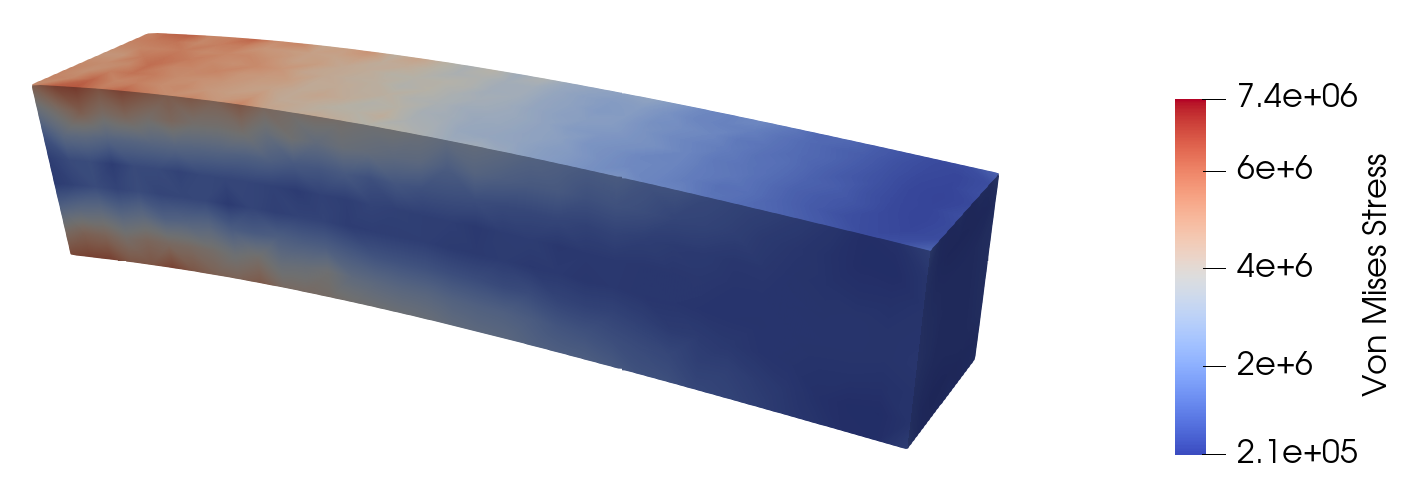}
	\caption{The contour of stress caluculated using juSFEM} \label{fig13}
\end{figure}

\begin{figure}[!h]
	\centering
	\includegraphics[width=0.9\textwidth]{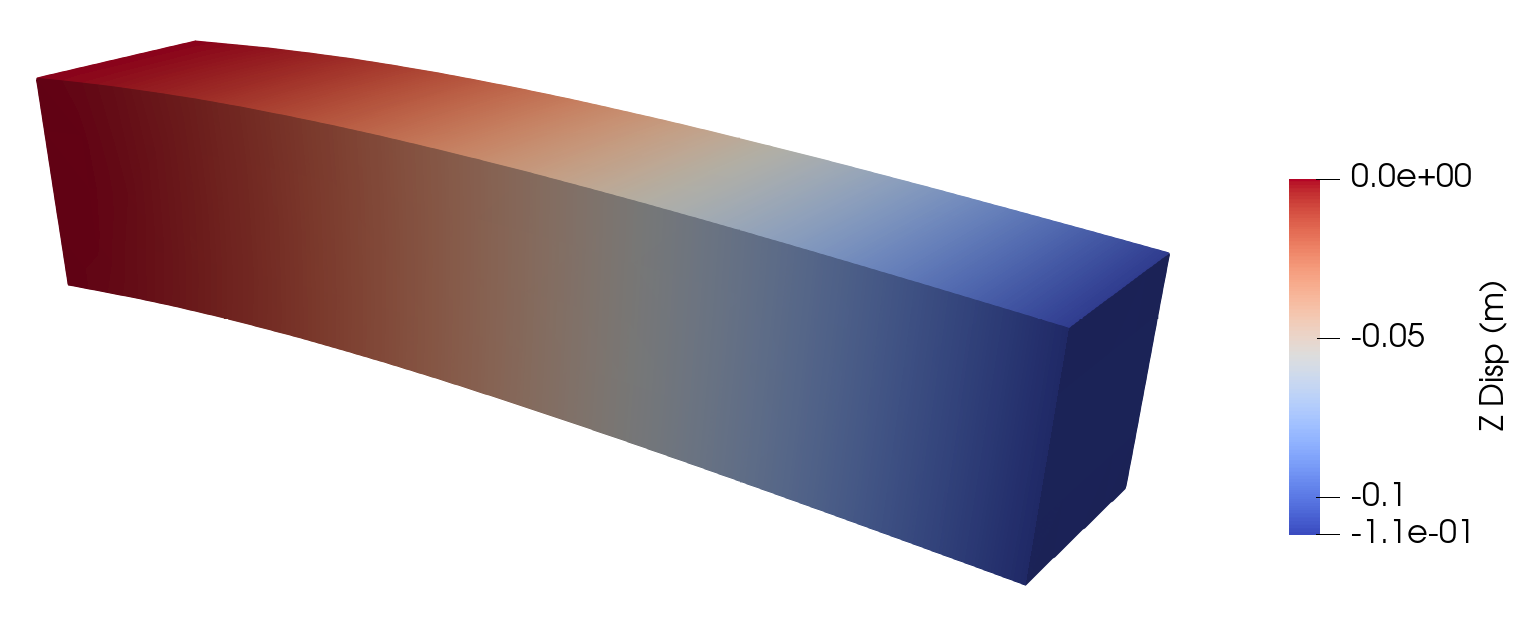}
	\caption{The contour of displacement caluculated using juSFEM} \label{fig14}
\end{figure}

(1)	The displacements and stress are calculated using juSFEM with the mesh model consisting of 458 nodes and 1,576 tetrahedral elements (T4 elements); see Figure \ref{fig13} and Figure \ref{fig14}.

(2)	The displacements are calculated using the conventional FEM with the mesh model consisting of 458 nodes and 1,576 tetrahedral elements (T4 elements).

(3)	According to reference \cite{liu2016smoothed}, the commercial software ABAQUS is used to calculate the displacements with a very fine mesh consisting of 45,779 nodes and 30,998 10-node tetrahedron elements (T10 elements).

\begin{figure}[H]
	\centering
	\includegraphics[width=1\textwidth]{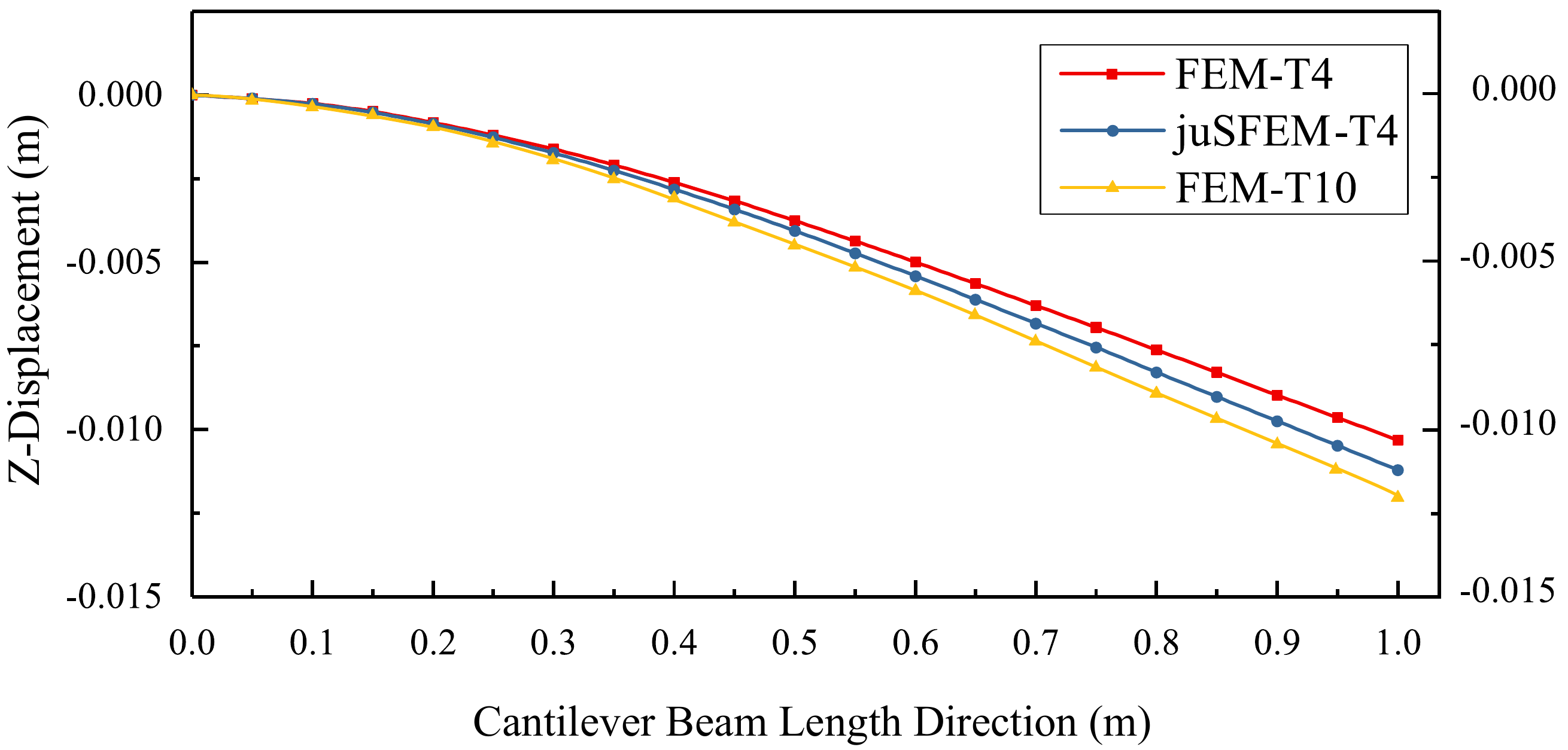}
	\caption{Comparison of the Cantilever beam deflection curves} \label{fig15}
\end{figure}

As shown in Figure \ref{fig15}, the displacements calculated using juSFEM is more accurate than that of the FEM-T4, and is a little less accurate than that of the FEM-T10. Thus, it can be considered that the correctness and accuracy of juSFEM are verified.

\subsection{Evaluation of the Efficiency of juSFEM}

The developed package juSFEM can be executed (1) in serial on a single CPU core or (2) in parallel on multiple CPU cores. The efficiency of juSFEM in the above two scenes is recorded and compared. Moreover, to better evaluate the computational efficiency of juSFEM, for the same size of the cantilever beam model illustrated in Figure \ref{fig12}, four mesh models are generated; see the details of the meshes in Table \ref{tab3}. For each of the mesh model, both the serial and the parallel calculation time is recorded and compared. 

\begin{table}[!h]
	\caption{Details of the employed four mesh models}
	\centering
		\begin{tabular}{|c|c|c|}
			\hline
			\textbf{Mesh model (T4)} & \textbf{Number of Nodes} & \textbf{Number of Elements} \\ \hline
			\textbf{1}               & 94,066                   & 503,963                     \\ \hline
			\textbf{2}               & 191,132                  & 1,037,691                   \\ \hline
			\textbf{3}               & 275,655                  & 1,498,126                   \\ \hline
			\textbf{4}               & 382,375                  & 2,076,930                   \\ \hline
		\end{tabular}
	\label{tab3}
\end{table}

The calculation process is divided into (1) the assembly of the global stiffness matrix and (2) the solution of the system of linear equations. The solution is performed by invoking the library PARDISO, and the computational efficiency of the solving is not discussed. Here only the computational efficiency of assembling the global stiffness matrix is evaluated by comparing. 

\begin{figure}[!h]
	\centering
	\includegraphics[width=0.8\textwidth]{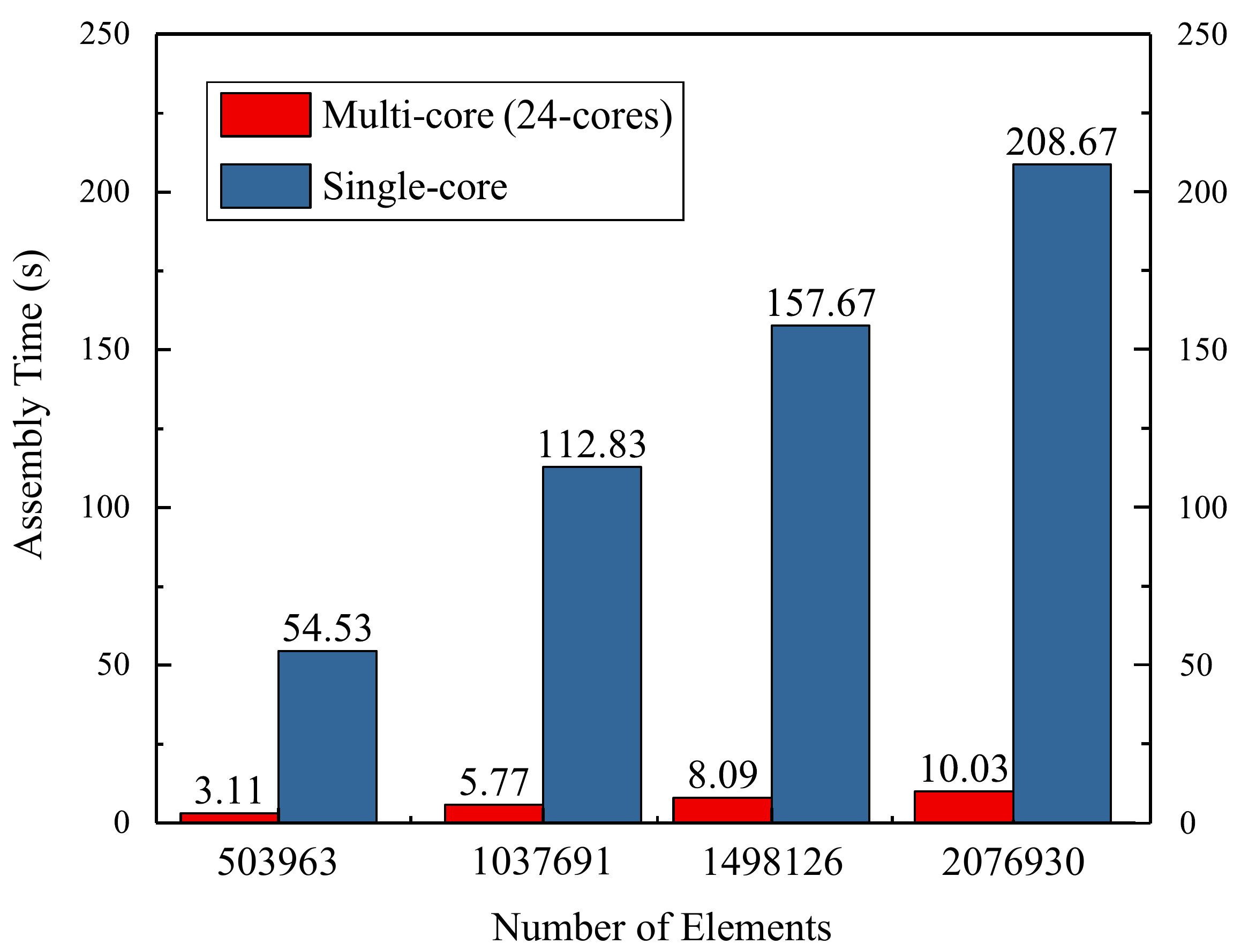}
	\caption{Comparison of the computational time for assembling the global stiffness matrix} \label{fig16}
\end{figure}

The calculation time for assembling the global stiffness matrix for each mesh model is compared in Figure \ref{fig16}. It can be seen from the above results that for the mesh model consisting of two million elements, it requires approximately 208s on the single-core, while it only needs approximately 10s on multiple cores. The speedup can reach 20$\times$ on the 24-cores CPU.

\begin{figure}[!h]
	\centering
	\includegraphics[width=0.8\textwidth]{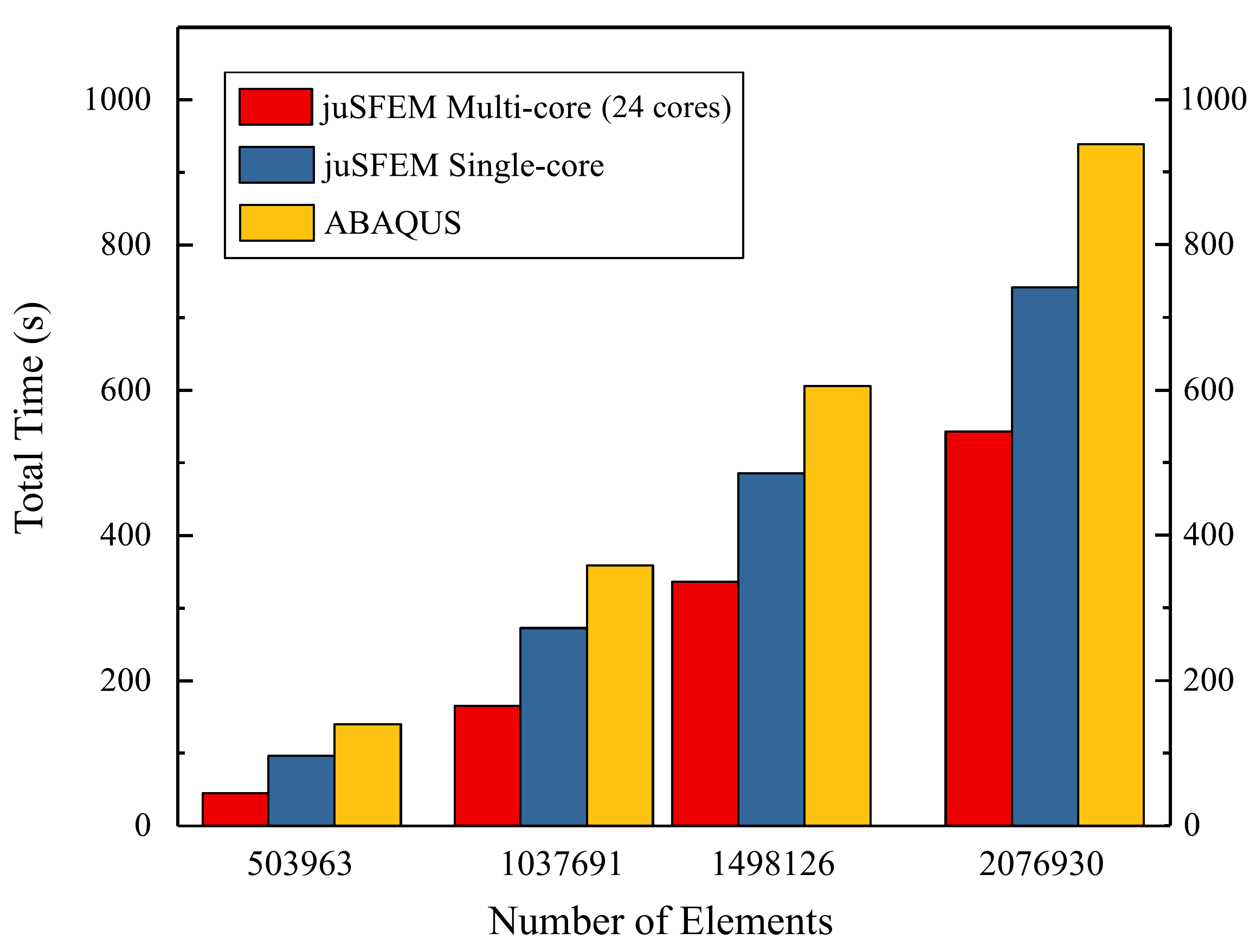}
	\caption{Comparison of the time spend on assembling stiffness matrix and solving the equation of the parallel juSFEM on (1) multi-core CPU, (2) the serial juSFEM on single-core CPU, and (3) the software ABAQUS} \label{fig17}
\end{figure}

To show the calculation efficiency of juSFEM, we also compared the times required by commercial software and juSFEM to calculate the same model; see the benchmark tests in Figure \ref{fig17}. In the comparisons, we recorded the total time of assembling the stiffness matrix and solving the equations. The results show that juSFEM is approximately 1.8$\times$ faster than ABAQUS.

\section{Discussion}
\label{sec5}

\subsection{Comprehensive Evaluation of juSFEM}

For a numerical modeling software or package, the correctness of the calculation results should be first guaranteed. Moreover, the computational efficiency should be satisfactory especially when solving large-scale problems. In this paper, we design and implement the open-source package of parallel S-FEM, juSFM. Here we present a comprehensive evaluation of juSFEM by analyzing the correctness and efficiency. 

\subsubsection{Computational Correctness}

Through the verification of the cantilever beam calculation example presented above, a comparison of the calculation error between the FEM and juSFEM is listed in Table \ref{tab4}.

\begin{table}[H]
	\caption{Verification of the correctness by comparing the calculated displacements}
	\centering
	\begin{tabular}{|c|c|c|c|}
		\hline
		\multirow{2}{*}{\textbf{Position}} & \multicolumn{3}{c|}{\textbf{Method}}                    \\ \cline{2-4} 
		& \textbf{FEM-T4} & \textbf{juSFEM-T4} & \textbf{FEM-T10} \\ \hline
		\textbf{0.2 m}                     & -8.20E-04 m     & -8.66E-04 m        & -9.54E-04 m      \\ \hline
		\textbf{0.4 m}                     & -2.61E-03 m     & -2.82E-03 m        & -3.08E-03 m      \\ \hline
		\textbf{0.6 m}                     & -4.99E-03 m     & -5.41E-03 m        & -5.86E-03 m      \\ \hline
		\textbf{0.8 m}                     & -7.62E-03 m     & -8.28E-03 m        & -8.92E-03 m      \\ \hline
		\textbf{1.0 m}                     & -1.03E-02 m     & -1.12E-02 m        & -1.20E-02 m      \\ \hline
	\end{tabular}
	\label{tab4}
\end{table}

When using numerical results calculated by the FEM-T10 as the baseline, the displacement error of juSFEM at the maximum deflection of the cantilever beam is 6.6\%, while that of the FEM T4 is 14.2\%. Table \ref{tab4} shows that the displacement variation of the FEM-T4 is the smallest, while the displacement solution of the juSFEM-T4 is more accurate. This is because the stiffness matrix of the FEM is too rigid for the same mesh, and only the upper bound of displacement solution can be obtained. The S-FEM optimizes the stiffness matrix based on the operation between elements, making the displacement solution closer to the baseline.

\subsubsection{Computational Efficiency}

\begin{figure}[H]
	\centering
	\includegraphics[width=0.8\textwidth]{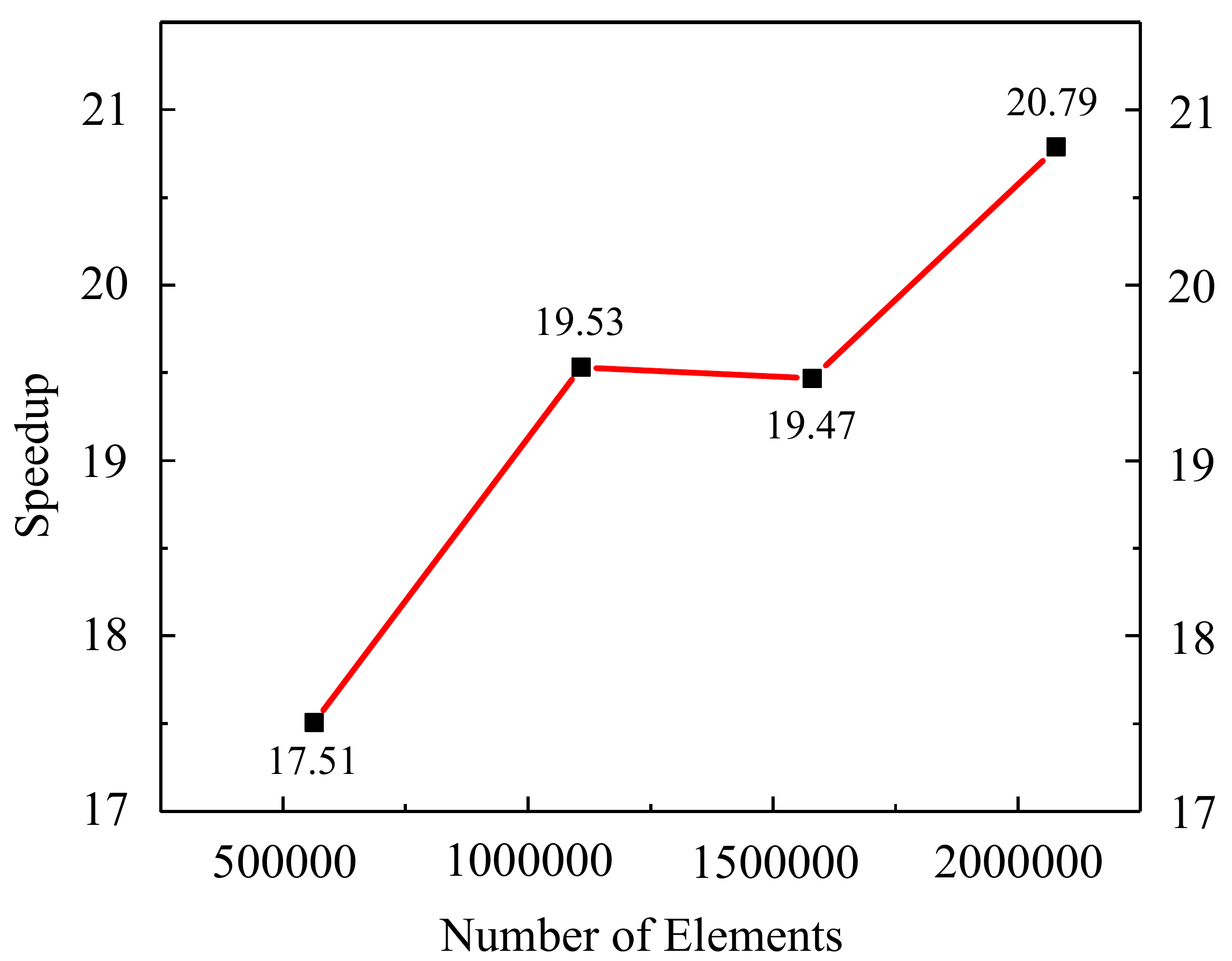}
	\caption{The speedup of assembling global stiffness matrix in parallel} \label{fig18}
\end{figure}

In the second verification example, we evaluated the computational efficiency of juSFEM by comparing the serial and parallel calculation time for four mesh models. The speedup of the parallel version over the corresponding serial version are illustrated in Figure \ref{fig18}.

As shown in Figure \ref{fig18}, for the mesh model composed of approximately 0.5 million elements, the speedup is 17.51$\times$ on a 24-core CPU, which is not close to the theoretical speedups of 24$\times$. This is because it requires additional time for task allocation in the multi-core parallelism. When the amount of computation is not large, the total computing time is too short, which leads to a larger proportion of the time consumed in task allocation. As the elements of computation increases, the time required to allocate tasks will increase slightly, but this part of the time is becoming shorter in the total computing time. For the mesh model composing of approximately 2 million elements, the efficiency of parallel computing on the multi-core CPU significantly improves, offsetting the time required for task allocation. In this case, the speedup reaches 20.79$\times$ and shows an upward trend. 

Another strategy to improve the computational efficiency is to reduce the memory operations by compressing the global stiffness matrix. In juSFEM, the COO format is used to assemble the global stiffness matrix; see the effect of compression of the global stiffness matrix in Table \ref{tab5}.

\begin{table}[H]
	\caption{Memory usage to store the global stiffness matrix}
	\centering
	\begin{tabular}{|c|c|c|c|c|}
	\hline
	\textbf{Dataset} & \textbf{\begin{tabular}[c]{@{}c@{}}Number of \\ Nodes\end{tabular}} & \textbf{\begin{tabular}[c]{@{}c@{}}Uncompressed\\ (GB)\end{tabular}} & \textbf{\begin{tabular}[c]{@{}c@{}}Compressed\\ (GB)\end{tabular}} & \textbf{\begin{tabular}[c]{@{}c@{}}Compression Ratio\\ (\%)\end{tabular}} \\ \hline
	\textbf{1}       & 94,066                                                              & 593.33                                                               & 4.21                                                                & 0.71                                                                      \\ \hline
	\textbf{2}       & 191,132                                                             & 2449.62                                                              & 8.27                                                                & 0.34                                                                      \\ \hline
	\textbf{3}       & 275,655                                                             & 5095.24                                                              & 11.77                                                               & 0.23                                                                      \\ \hline
	\textbf{4}       & 382,375                                                             & 9804.19                                                              & 15.48                                                               & 0.16                                                                      \\ \hline
	\end{tabular}
	\label{tab5}
\end{table}

For the large-scale mesh model, there are many zero elements in the global stiffness matrix. Storing the global stiffness matrix in the form of the full matrix is quite memory-consuming. When the number of nodes reaches 382,375, it takes approximately 1 TB of memory to store the global stiffness matrix. Due to hardware specifications, it will be very difficult to compute the large-scale mesh model. juSFEM adopts the COO compression format and stores only non-zero elements in the global stiffness matrix, which significantly reduces memory usage. After assembling the stiffness matrix using the COO format, the stiffness matrix is then converted to the CSC format to further reduce the memory requirements, and used to generate the system of linear equations. The solving of the system of the linear equations becomes more efficient when using the CSC format to store the coefficient matrix and global stiffness matrix. 

As mentioned above, the computation time of juSFEM is divided into two parts: (1) assembling the global stiffness matrix and (2) solving the equations. In order to analyze the performance of juSFEM, a mesh model composing of two million tetrahedral elements was selected. We recorded the proportion of these two parts in the total computing time in parallel and serial cases respectively.

\begin{figure}[H]
	\centering
	\includegraphics[width=1\textwidth]{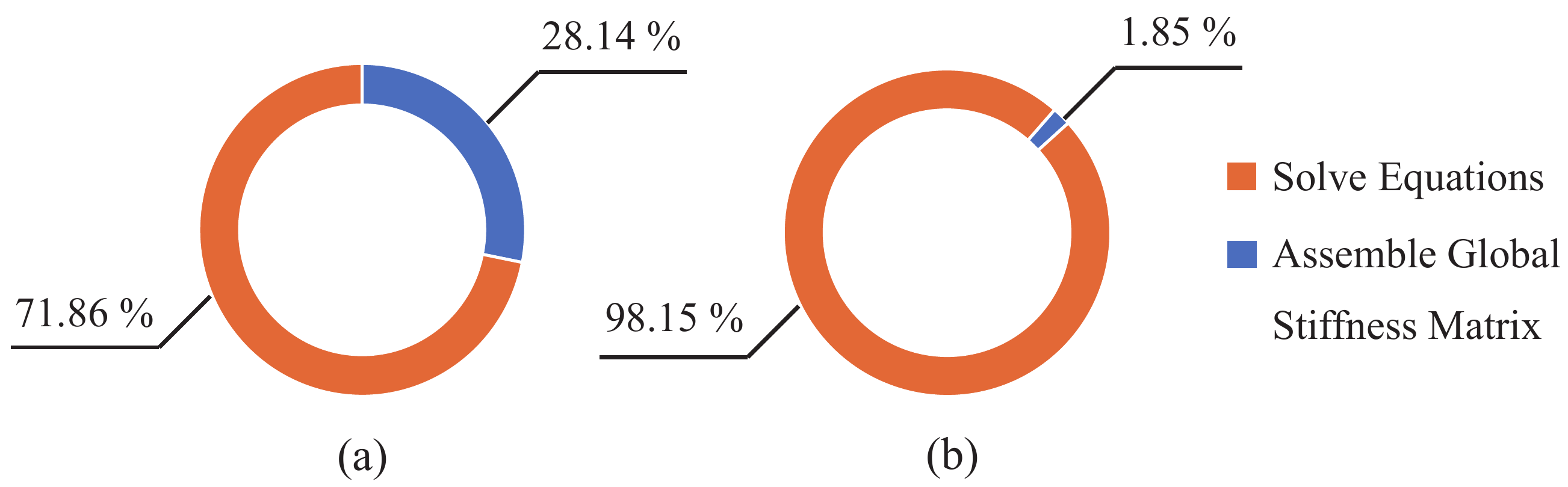}
	\caption{The proportion of the time spent on assembling stiffness matrix and solving the equation ((a) single-core (b)multi-core)} \label{fig19}
\end{figure}

As shown in Figure \ref{fig19}, when the parallel juSFEM is used for the mesh model composing of approximately 2 million tetrahedron elements, the time of assembling the global stiffness matrix is only 1.85\% of the total calculation time, and the rest of the time is spent on solving equations. Thus, the computationally most expensive stage in juSFEM is to solve the system of equations. Open-source libraries such as the BLAS, Sparse BLAS, LAPACK, and MKL \cite{burylov2007intel} are commonly used to solve the equations. However, reference \cite{schenk2004solving} shows that PARDISO reaches a high-speed in solving the equations. 

In summary, juSFEM realizes multi-core parallel high-performance S-FEM elastic problem by redesigning parallel algorithm and combining the advantages of Julia language. Although juSFEM can make full use of the multi-core processor, it takes a lot of time to solve the linear equations of large sparse matrix. Moreover, the number of cores in a multi-core computer will also limit the speedup of juSFEM.

\subsection{Comparison with Other S-FEM Programs}

At present, there are very few studies on the S-FEM programs, and most of them use programming languages such as C/C++ or Fortran. However, to achieve efficient computing, high demand of programming skills is required. In recent years, MATLAB and Python become popular. Although their code is more readable, but in general it is quite computationally expensive \cite{nickolls2008scalable}.

juSFEM has a clear structure, all callers are constructed in a modular way, and each calculation step has a high degree of customization, which is characterized by high efficiency and brevity. Meanwhile, juSFEM has competitive computational efficiency.

In reference \cite{li2016automatic}, a set of connectivity lists was developed for the S-FEM. These connectivity lists will be created simultaneously, which will reduce computation and time consumption. In addition, the centroid of the mesh is selected as a uniform index to avoid duplicate storage.

In juSFEM, we choose the FS-FEM. The face is used as the uniform index in the program. However, we redesign the method of storing mesh detail, which not only avoid the problem of repeated storage but also considered parallel computation. juSFEM stores the global stiffness matrix directly in a compressed fashion, which saves a lot of memory and facilitates large-scale mesh computing.

As discussed above, juSFEM is computationally accurate and efficient. Under the redesigned calculation framework, the power of parallelism is well utilized. Meanwhile, the calculation process in juSFEM is packaged in the form of a function with concise grammar, which makes it convenient for further improvement. 

\subsection{Outlook and Future Work}

Currently, the S-FEM is widely used in material science and mechanical science. We hope that juSFEM based on the S-FEM can be commonly used in geomechanics to provide fast and reliable numerical modeling results, for example, to analyze the slope deformation and failure \cite{8895751}.

We also plan to extend juSFEM to analyze the plasticity problems. In the plasticity problems, we need to solve the equations iteratively. In regard to large-scale computational models, the solution time of equations becomes particularly important \cite{nyssen1981efficient}. We plan to adopt a more efficient parallel method to accelerate the solution speed of the equations on the GPU \cite{garland2008parallel,ge2013multi}. 

In juSFEM, we use two methods to plot the results in post-processing. ParaView needs to import the mesh file rewritten by juSFEM, which is very tedious. When using the PlotlyJS, it can only plot the displacements of the model in juSFEM, and thus needs to be improved in the future. We hope in juSFEM the post-processing is conducted in a convenient and efficient way.

Liu G.R. et al. \cite{liu2016smoothed} proposed the S-FEM by combining the gradient smoothing technique with the FEM, and the computation results are closer to the analytical solution. But in the S-FEM, there is the problem of element incongruity in higher-order interpolation. To further improve the accuracy of interpolation, Liu G.R. and Zhang \cite{gui2013smoothed} proposed a smoothed point interpolation method (S-PIM) based on $G$ space theory and double weak forms \cite{liu2010ag} by developing generalized gradient smoothing technology. The S-PIM can reduce the error by optimizing the interpolation form. We also hope to use Julia language to implement the parallel S-PIM which is an extension of juSFEM.

\section{Conclusion}
\label{sec6}

In this paper, we have designed and implemented an open-source package of the parallel S-FEM for elastic problems by utilizing the Julia language on multi-core CPU. We specifically designed a new strategy for the computation of the element stiffness matrices in parallel. The designed strategy is well suitable to be parallelized on the multi-core CPU. To verify the correctness and evaluate the efficiency of juSFEM, two groups of benchmark tests have been conducted. We have found that (1) juSFEM only required 543 seconds to calculate the displacements of a 3D elastic cantilever beam model which was composed of approximately 2 million tetrahedral elements, while in contrast, the commercial FEM software required 930 seconds for the same calculation model; (2) the parallel juSFEM executed on the 24-core CPU is approximately 20$\times$ faster than the corresponding serial version; (3) juSFEM employed COO compression format to store the global stiffness matrix, and the compression ratio can achieve to be 0.16\% for the mesh model composed of approximately 2 million tetrahedrons. juSFEM is computationally efficient and can achieve accurate numerical results. Moreover, the structure and function of the juSFEM are easily modularized, and the Julia code in juSFEM is clear and readable, which is convenient for further development.

\section*{Acknowledgments}
This research was jointly supported by the National Natural Science 
Foundation of China (Grant Numbers: 11602235 and 41772326), and the Fundamental Research Funds for China Central Universities (Grant Numbers: 2652018091, 2652018107, and 2652018109). The authors would like to thank the editor and the reviewers for their contributions.

  \bibliographystyle{elsarticle-num} 
  \bibliography{reflatex}
 

\end{document}